\documentclass[10pt,a4paper]{article}
\usepackage{float}
\usepackage{graphicx}
\usepackage{amsmath}
\usepackage{amssymb}
\usepackage{latexsym}
\usepackage{color}

\usepackage{cite}
\usepackage{caption}
\usepackage{subcaption}
\usepackage{refstyle}
\usepackage{lipsum}

\providecommand{\openone}{\leavevmode\hbox{\small1\kern-4.3pt\normalsize1}}

\begin{document}

\bigskip \thispagestyle{empty}

\begin{center}
\vspace{1.8cm}

\textbf{\Large Quantum Phase Sensitivity with Generalized Coherent States Based on Deformed $su(1,1)$ and Heisenberg Algebras}


\vspace{1.5cm}

{\bf Nour-Eddine Abouelkhir
}$^{a}${\footnote {email: {\sf abouelkhir115@gmail.com}}}, {\bf Abdallah Slaoui }$^{a,b,c}${\footnote {email: {\sf abdallah.slaoui@um5s.net.ma}}} and {\bf Rachid Ahl Laamara}$^{a,b}${\footnote {email: {\sf r.ahllaamara@um5r.ac.ma}}}

\vspace{0.5cm}

$^{a}${\it LPHE-Modeling and Simulation, Faculty  of Sciences,
	Mohammed V University in Rabat, Rabat, Morocco.}\\
$^{b}${\it Centre of Physics and Mathematics, CPM, Faculty  of Sciences, Mohammed V University in Rabat, Morocco.}\\
$^{c}${\it Center of Excellence in Quantum and Intelligent Computing, Prince Sultan University, Riyadh, Saudi Arabia.}\\[1em]

\vspace{1.60cm} \textbf{Abstract}
\end{center}

\baselineskip=18pt \medskip
We investigate the phase sensitivity of a Mach–Zehnder interferometer using a special class of generalized coherent states constructed from generalized Heisenberg and deformed $su(1,1)$ algebras. These states, derived from a perturbed harmonic oscillator with a four-parameter deformed spectrum, provide enhanced tunability and nonclassical features. The quantum Fisher information and its associated quantum Cramér-Rao bound are computed to define the fundamental precision limits in phase estimation. We analyze the phase sensitivity under three realistic detection methods: difference intensity detection, single-mode intensity detection, and balanced homodyne detection. The performance of each method is compared with the quantum Cramér–Rao bound to evaluate their optimality. Our results demonstrate that, for suitable parameter regimes, these generalized coherent states enable phase sensitivities approaching the quantum limit. This offers a flexible framework for precision quantum metrology and potential applications in quantum-enhanced sensing.

\vspace{0.25cm}

\section{Introduction}
Interferometry is a technique based on the interference of waves. This technique is essential for precision measurement, quantum metrology and sensing, and is important for developing fundamental physics concepts \cite{Abbott2017}. Interferometric measurements of physical quantities like magnetic and gravitational fields fundamentally depend on detecting phase shifts in the interfering waves \cite{Ou2020}. Due to their exceptional sensitivity to phase changes, interferometers are extensively utilised in precision measurement and metrology. Specifically, quantum metrology is gaining widespread application across diverse fields, driven by advances in quantum technologies \cite{Liu2017,Dowling2015} and quantum information theory \cite{Gerry2012, Gagatsos2013, Gaidi2024, Alaoui2024, Gaidi2025}. As a fundamental aspect of quantum metrology, phase estimation has received considerable research attention. The sensitivity of optical interferometry is a research topic of interest and plays an essential role in various precision metrology applications \cite{Helstrom1969,Slaoui2023,Slaoui2022,Sparaciari2015,Sparaciari2016}, such as environmental sensing, gravitational wave detection \cite{Grote2013, Vahlbruch2018, Tse2019}, quantum technologies \cite{Giovannetti2012, Lang2014}, and gyroscopes \cite{Khial2018}.\par

When using classical resources in conventional interferometry, measurement sensitivity is fundamentally bounded by the shot-noise limit (SNL), scaling as $1/\sqrt{N}$, where $N$ represents the total photon number \cite{Caves1981, Giovannetti2011}. This boundary is alternatively termed the standard quantum limit (SQL). Caves \cite{Caves1981} introduced the use of squeezed states as a strategy to surpass the SQL. The method accounts for vacuum fluctuations coupled through the interferometer's unpopulated input mode. Subsequently, various quantum resources \cite{Xiao1987, Steinlechner2013} have since been employed to enhance measurement precision. Quantum resources such as entangled coherent states, NOON states \cite{Bollinger1996, Dowling2008}, two-mode squeezed states \cite{Anisimov2010}, and number-squeezed states \cite{Pezzé2013} have been employed to enhance phase estimation precision, offering the potential to reach the Heisenberg limit (HL), where the precision scales as $1/N$ \cite{Ou1997, Giovannetti2006, Anisimov2010}. In quantum interferometry, the theoretical bounds on phase sensitivity are obtained by applying the quantum Fisher information (QFI), along with its associated quantum Cramér-Rao bound (QCRB) \cite{Braunstein1994, Pezzé2015}. These bounds are not only of theoretical interest, but also extremely useful for evaluating the optimality of realistic detection schemes.\par

The SU(1,1) interferometer has been introduced as a novel variant of the traditional Mach–Zehnder interferometer (MZI). Yurke et al. \cite{Yurke1986} introduced it theoretically, replacing linear beam splitters (BS) with non-linear beam splitters to split and mix two input fields coherently, achieving precise phase estimation. The interferometer is named SU(1,1) because it uses a specific type of nonlinear interaction that arises in parametric processes following SU(1,1) symmetry. This kind of interaction is not the same as the SU(2) interaction linked to linear wave mixing by a beam splitter. This nomenclature is related to the specific nature of the interaction used in these two types of interferometers. In previous work \cite{Abdellaoui2024, Abouelkhir2025}, we demonstrated that SU(2) spin-coherent states and SU(1,1) coherent states can surpass classical limits under optimized detection schemes.\par

In this work, we investigate phase sensitivity in an MZI configuration. It is well known that a wide class of linear interferometric setups can be effectively recast into an MZI framework. This allows for the systematic optimisation of phase estimation strategies, depending on the chosen input state and detection schemes. However, this equivalence does not extend to nonlinear configurations, such as the SU(1,1) interferometer.\par

Achieving optimal phase sensitivity in an interferometric setup requires careful consideration of both the input state and the detection scheme. The QFI is central to this endeavor, as it establishes the ultimate precision bound via the QCRB, given by $\Delta\phi_{\mathrm{QCRB}} = 1/\sqrt{F}$ \cite{Helstrom1973, Helstrom1968, Holevo1973, Wu2019, Abouelkhir2023}. Consequently, maximizing the QFI is a key objective of quantum metrology. In practice, however, the phase sensitivity $\Delta\phi_{\mathrm{det}}$ achieved by a specific detection scheme is generally bounded from below by the QCRB, i.e., $\Delta\phi_{\mathrm{det}} \geq \Delta\phi_{\mathrm{QCRB}}$. Therefore, identifying detection strategies that closely approach this bound is crucial for implementing high-precision quantum measurements.\par

In this work, we consider a special class of generalized coherent states based on the algebraic structures of the generalized Heisenberg algebra (GHA) and the generalized $su(1,1)$ algebra. These coherent states are constructed for a perturbed harmonic oscillator with a four-parameter deformed energy spectrum [1, 2]. In particular, the focus is on the generalized $su(1,1)$ coherent states, which extend the standard SU(1,1) Perelomov \cite{Perelomov1977} and Barut-Girardello \cite{Barut1971} constructions by incorporating a nonlinear deformation through the characteristic function $\varphi(\mathcal{H})$. These states have attracted interest due to their rich algebraic structure and potential nonclassical features [1, 5, 6]. Despite the extensive utilisation of the standard SU(1,1) coherent states in the domains of quantum optics and interferometry [3, 7], the deformed versions provide enhanced flexibility and tunable properties. In the context of quantum interferometry, the utilisation of these generalized SU(1,1) coherent states as input states facilitates enhanced control of phase sensitivity, with the potential to achieve optimised performance under realistic conditions. The coherent states presented here provide a foundation for future investigation of their quantum statistical properties, squeezing, and robustness under decoherence [8, 9].\par

The structure of this paper is as follows. In Section II, we present the construction of generalized coherent states based on the GHA and the generalized $su(1,1)$ algebra, including the derivation of coherent states for a perturbed harmonic oscillator. Section III is devoted to the QFI and phase estimation using the generalized coherent states within an MZI. In Section IV, we analyze phase sensitivity using three realistic detection schemes difference intensity detection, single mode intensity detection, and balanced homodyne detection under various interferometric scenarios. Finally, we conclude in Section V with a summary of our findings and a discussion on potential future directions.

\section{Construction of coherent states} 
\subsection{Generalised Heisenberg algebra}
We start with a short introduction to the GHA. This algebra consists of three generators, namely $\mathcal{H}$, $\mathcal{B}^{\dagger}$, and $\mathcal{B}$, which satisfy the following relations:
\begin{align} 
\mathcal{H} \mathcal{B}^{\dagger} = \mathcal{B}^{\dagger} \varphi(\mathcal{H}), \hspace{1cm}
\mathcal{B} \mathcal{H} = \varphi(\mathcal{H}) \mathcal{B}, \label{AH}
\end{align}
and the commutation
\begin{align}\label{AA+}
[\mathcal{B}, \mathcal{B}^{\dagger}] = \varphi(\mathcal{H}) - \mathcal{H}.
\end{align}

Here, $\varphi(\mathcal{H})$ is an analytic function of $\mathcal{H}$, referred to as the characteristic function of the algebra. The commutator is defined as $[\mathcal{B},\mathcal{B}^{\dagger}] = \mathcal{B}\mathcal{B}^{\dagger} - \mathcal{B}^{\dagger}\mathcal{B}$, with the adjoint relations $(\mathcal{B}^{\dagger})^{\dagger} = \mathcal{B}$ and $\mathcal{H}^{\dagger} = \mathcal{H}$, where $\mathcal{H}$ denotes the dimensionless Hamiltonian of the system under consideration. By appropriately selecting the function $\varphi(\mathcal{H})$, one can generate a broad class of Heisenberg-type algebras. Specifically, when $\varphi(\mathcal{H}) = \mathcal{H} + 1$, the generalized Heisenberg algebra (GHA), as defined by equations (\ref{AH}), and (\ref{AA+}), reduces to the standard Heisenberg algebra, which is generated by the dimensionless Hamiltonian $H$ along with the harmonic oscillator's creation and annihilation operators, $c^{\dagger}$ and $c$. The Casimir operator associated with the GHA is given by
\begin{equation}
    C=B^{\dagger}B-\mathcal{H}=BB^{\dagger}-\varphi(\mathcal{H}).
\end{equation}
For an eigenvector $|m\rangle$ of the Hamiltonian H, the irreducible representation of the GHA is expressed as follows
\begin{align}
&\mathcal{H}|m\rangle=\epsilon_{m}|m\rangle,\\
&B^{\dagger}|m\rangle=N_{m}|m+1\rangle,\\
&B|m\rangle=N_{m-1}|m-1\rangle,
\end{align}
where 
\begin{equation} \label{N GHA}
    N_{m}^{2}=\epsilon_{m+1}-\epsilon_{0},
\end{equation}
and $\epsilon_{m+1}$ and $\epsilon_{m}$ denote two successive eigenvalues of Hamiltonian $\mathcal{H}$ such that $\epsilon_{m+1}=\varphi(\epsilon_{m})$ and $\epsilon_{0}$ represents the eigenvalue associated with the vacuum state. The operators $B$ and $B^{\dagger}$ are identified as the annihilation and creation operators of the GHA, respectively, with the vacuum condition $B|0\rangle = 0$ being satisfied. Consequently, the eigenvalue $\varepsilon_n$ corresponds to the $n$-th iteration of $\varepsilon_0$ under the function $\varphi$, that is, $\varepsilon_n = \varphi^{n}(\varepsilon_0)$. For the particular function $\varphi(\mathcal{H}) = q\mathcal{H} +1$, the equations (\ref{AH}), and (\ref{AA+}) recover the q-harmonic oscillator algebra. In the limit $q\rightarrow 1$, this algebra becomes the standard harmonic oscillator algebra. Moreover, the GHA has already been shown to be applicable to several physical systems characterized by a known spectrum, such as the Morse potential \cite{Curado2008}, the square well potential \cite{Hussin2011}, and the Poschl-Teller potential \cite{Bagarello2018}.

\subsection{Generalised su(1,1) algebra}
Analogous to the GHA, we now introduce an algebraic framework that extends the conventional su(1,1) algebra, along with its associated Fock space representation. The generalized su(1,1) algebra is characterized by three operators: the Hamiltonian $\mathcal{H}$ and the ladder operators $G_+$ and $G_-$, which satisfy the following relations
\begin{align} 
\mathcal{H}\mathcal{H}_{+}=&G_{+} \varphi(\mathcal{H}),\\ 
G_{-}\mathcal{H}=&\varphi(\mathcal{H})G_{-},
\end{align}
and the commutation
\begin{align}
[G_{+},G_{-}]=&(\mathcal{H}-\varphi(\mathcal{H}))(\mathcal{H}+\varphi(\mathcal{H})-1),
\end{align}
where  $(G_{+})^{\dagger}=G_{-}$, $\varphi(\mathcal{H})$ is an analytic function of $\mathcal{H}$, and $\mathcal{H}^{\dagger}=\mathcal{H}$ is the dimensionless Hamiltonian of the system considered. In fact, $\mathcal{H}$ can be any operator that is hermitian. The Casimir operator in this algebra is given by
\begin{equation}
    C=G_{+}G_{-}-\mathcal{H}(\mathcal{H}-1)=G_{-}G_{+}-\varphi(\mathcal{H})(\varphi(\mathcal{H})-1).
\end{equation}
We now provide the Fock space representation of the generalized su(1,1) algebra. For an eigenstate $|m\rangle$ of the Hamiltonian $\mathcal{H}$, the irreducible action of the algebra's generators is described by the following relations:
\begin{align}
&\mathcal{H}\left| m\right\rangle =\epsilon_{m}\left| m\right\rangle ,\\
&G_{+}\left| m\right\rangle =\mathcal{N}_{m}\left|m+1\right\rangle,\\
&G_{-}\left|m\right\rangle=\mathcal{N}_{m-1}\left|m-1\right\rangle,
\end{align}
where 
\begin{equation} \label{N su}
    \mathcal{N}_{m}^{2}=(\epsilon_{m+1}-\epsilon_{0})(\epsilon_{m+1}+\epsilon_{0}-1).
\end{equation}
Similarly to the GHA, the operators $G_{+}$ and $G_{-}$ are interpreted as the annihilation and creation operators of the generalized su(1,1) algebra, respectively, and the vacuum state condition $G_{-}|0\rangle=0$ is verified.

\subsection{Construction of coherent states}
In fact, coherent states are seen as the closest quantum analogue to classical radiation fields, reproducing many of the wave properties of light \cite{Zhang1990}. Introduced by Glauber \cite{Glauber1963}, they have acquired a central place in quantum optics, particularly after Glauber showed them to be the eigenstates of the harmonic oscillator's annihilation operator while simultaneously realizing the minimal bound of the Heisenberg uncertainty relation \cite{Dur2000, Kirdi2023,Bakraoui2022}. Subsequently, Perelomov generalized Glauber's approach to the framework of arbitrary Lie groups \cite{Perelomov1977,Sanders2012}. In this formalism, a generalized coherent state is obtained simply by applying a generalized displacement operator defined as the unitary exponentiation of the considered Lie algebra to an extremal state. In this subsection, we construct the coherent states corresponding to the GHA and the generalized su(1,1) algebra. We begin with the GHA coherent state, defined as the eigenstate of the annihilation operator
\begin{equation} \label{Az}
    B|\zeta\rangle=\zeta|\zeta\rangle,
\end{equation}
where the eigenvalue $\zeta$ may take complex values due to the non-Hermitian nature of the operator $B$. The state $|\zeta\rangle$ can be expressed as a superposition of number states $|m\rangle$:
\begin{equation} \label{z}
    |\zeta\rangle=\sum_{m=0}^{\infty}\alpha_{m}|m\rangle,
\end{equation}
where $\alpha_{m}$ denotes complex coefficients satisfying $\sum_{m=0}^{\infty}|\alpha_{m}|^{2}=1$.
Using Equations (\ref{Az}) and (\ref{z}), the action of the annihilation operator $B$ on the state $|\zeta\rangle$ is given by
\begin{equation} 
    B|\zeta\rangle=\sum_{m=0}^{\infty}\alpha_{m+1}N_{m}|m\rangle=\zeta\sum_{m=0}^{\infty}\zeta_{n}|m\rangle.
\end{equation}
The solution of the equation $\alpha_{m+1}N_{m}=\zeta \alpha_{m}$ for the coefficients of $\left|m\right\rangle$ is
\begin{equation}
    \alpha_{m}=\frac{\alpha_{0}\zeta^{n}}{N_{m-1}!},
\end{equation}
where $N_{m}!=N_{m}N_{m-1}...N_{0}$, and by consistency $N_{-1}!=1$. Therefore, the state $|\zeta\rangle$ can be expressed as
\begin{equation} \label{z GHA}
    |\zeta\rangle=N(|\zeta|)\sum_{m=0}^{\infty}\frac{\zeta^{n}}{N_{m-1}!}|m\rangle,
\end{equation}
where $N(|\zeta|)=\alpha_{0}$. To find the normalization factor $N(|m|)$, we use the condition $\langle \zeta|\zeta\rangle=1$, yielding
\begin{equation} \label{Norma GHA}
    N(|\zeta|)=\left(\sum_{m=0}^{\infty}\frac{|\zeta|^{2m}}{(N_{m-1}!)^{2}}\right)^{-1/2}.
\end{equation}
To construct the coherent states related to the generalized su(1,1) algebra, one simply substitutes $N_{m}$ with $\mathcal{N}_{m}$ in the expression of $|\zeta\rangle$ (Eq. \ref{z GHA}) and in the normalization factor $N(|\zeta|)$ (Eq. \ref{Norma GHA}) 
\begin{equation} \label{z su}
    |\zeta\rangle=\mathcal{N}(|\zeta|)\sum_{m=0}^{\infty}\frac{\zeta^{n}}{\mathcal{N}_{m-1}!}|m\rangle,
\end{equation}
where
\begin{equation} \label{Norma su}
    \mathcal{N}(|\zeta|)=\left(\sum_{m=0}^{\infty}\frac{|\zeta|^{2m}}{(\mathcal{N}_{m-1}!)^{2}}\right)^{-1/2}.
\end{equation}
The deformation parameters encoded in $N_m$ or $\mathcal{N}_m$ modify the oscillator’s energy-level spacing, introducing nonlinearities into the system. This allows interpolation between standard Glauber coherent states describing classical like light fields and deformed states with enhanced nonclassical properties such as squeezing and sub-Poissonian statistics. Glauber originally introduced coherent states to represent fields with Poissonian photon number statistics and minimal uncertainty \cite{Glauber1963}. Similar nonclassical features have been studied in hypergeometric states \cite{FuSasaki1997}. In the limit where deformation vanishes, our formalism reduces to the standard harmonic oscillator coherent states, preserving the connection to quantum optics.

\subsection{Coherent State Construction under Harmonic Oscillator Perturbations}
In this context, we examine a modified dimensionless energy spectrum of the harmonic oscillator, expressed as
\begin{equation} \label{n}
    \varepsilon=m+\beta(n).
\end{equation}
The function $\beta(n)$ is given by
\begin{equation}
    \beta(n)=\frac{am+e}{kn+d}, \quad n\geq 0,
\end{equation}
where $a$, $k$, $d$, and $e$ are real and non-zero constants satisfying the following conditions \cite{Curado2013}
\begin{align}
&|a/k|<1,\\
&d/k>0,\\
&-\frac{4ad-4ke}{k^2}\geq r-1,\quad r\in[0,1].
\end{align}
For simplicity, we choose $k=1$.\\\\
For the GHA, coherent states (\ref{z GHA}) can be obtained by substituting Eq. (\ref{n}) into both Eq. (\ref{N GHA}) and Eq. (\ref{Norma GHA})
\begin{equation}
    N_{m-1}^{2}=\frac{m^2+(a+d)m+e}{m+d}-\frac{e}{d},
\end{equation}
and
\begin{equation}
     N(|\zeta|)=\left(\sum_{m=0}^{\infty}\frac{(m+d)!}{(m+d+a-e/d)!}\frac{|\zeta|^{2m}}{m!}\right)^{-1/2}.
\end{equation}
We can rewrite the above equation as
\begin{equation}
     N(|\zeta|)=\left[{}_{1}F_{1}(d+1,d+a-e/d+1;|\zeta|^{2})\right]^{-1/2},
\end{equation}
where \( {}_1F_1 \) is a hypergeometric function defined as
\begin{equation}
    {}_1F_1(\omega;\sigma;x) = \sum_{m=0}^{\infty} \frac{\omega(\omega+1)...(\omega+n-1)}{\sigma(\sigma+1)...(\sigma+n-1)} \frac{x^n}{m!}.
\end{equation}
Analogous to the GHA case, the coherent states in (\ref{z su}) corresponding to the generalized su(1,1) algebra are derived by inserting Eq. (\ref{n}) into Eqs. (\ref{N su}) and (\ref{Norma su})
{\small
\begin{equation}
    (\mathcal{N}_{m-1}!)^{2}=\frac{d(ad+2de+e)\Gamma(m+d+a-e/d+1)}{\Gamma(1+d+m)^{2}\Gamma(d+a-e/d+1)}\gamma\Gamma(d)^{2}m!
    ,
\end{equation}
}
where
{\small
\begin{align}
\gamma=&\left(\Gamma(\omega+m)\Gamma(\sigma+n)\right)/\left(\Gamma(\omega+1)\Gamma(\sigma+1)\right),\\
\omega=&\frac{1}{2}(a+d+1+\frac{e}{d})-\frac{1}{2}\sqrt{(a+d+1+\frac{e}{d})^{2}-4(a+2e+\frac{e}{d})},\\
\sigma=&\frac{1}{2}(a+d+1+\frac{e}{d})+\frac{1}{2}\sqrt{(a+d+1+\frac{e}{d})^{2}-4(a+2e+\frac{e}{d})},
\end{align}
}
and for the normalization factor, we have
\begin{equation}
    \mathcal{N}(|\zeta|)=\left[{}_{2}F_{3}(d+1,d+a;\omega,\sigma,a-e/d+d+1;|\zeta|^{2})\right]^{-1/2},
\end{equation}
where ${}_{2}F_{3}$ is the generalized hypergeometric function. The limit of the deformed harmonic oscillator is recovered when $a=1/2$ and $e=d/2$, i.e., under these conditions, it becomes the ordinary harmonic oscillator. In this case, the GHA coherent state $(\ref{z GHA})$ becomes the ordinary coherent state given by
\begin{equation} 
    |\zeta\rangle=\exp\left[{-\frac{|\zeta|^{2}}{2}}\right]\sum_{m=0}^{\infty}\frac{\zeta^{n}}{\sqrt{n!}}|m\rangle.
\end{equation}
The parameters $a$, $d$, and $e$ determine how strongly the oscillator departs from the harmonic potential, thereby controlling squeezing, photon-number fluctuations, and the QFI relevant to phase sensitivity. Larger nonlinearities enhance quantum features useful for metrology, while the ordinary oscillator limit recovers coherent states with Poissonian statistics \cite{Curado2013, Manko1997}.

\section{QFI and Phase Estimation with Generalized Coherent States in MZI}
The interferometric setup we are examining in this paper is the standard Mach-Zehnder interferometric setup, as depicted in Figure 1, with BS1 and BS2 denoting the two beam splitters, characterized by transmission (reflection) coefficients $\tau$ ($r$) and $\tau'$ ($r'$), respectively. The entire setup is divided into three parts. The first part is for the preparation of the light state that will be the input of the interferometer (second part), and the third part is for the detection. The initial input state is denoted as $|\psi_{in}\rangle$. After the first beamsplitter BS1, the state is represented as $|\psi\rangle$. Throughout this study, the input state is considered to be pure and unaffected by losses. In general, the precision of phase estimation is determined by the QFI, which depends on how the interferometer phase delay is modeled, including: (a) two independent phase shifts denoted as $\phi_{1}$ and $\phi_{2}$, respectively, where $\phi_{1}$ is in the upper arm and $\phi_{2}$ is in the lower arm of the interferometer; (b) a single phase shift only in the lower arm; and (c) a symmetrically distributed phase shift.

\begin{figure}[h] 
\centering
		\includegraphics[scale=0.25]{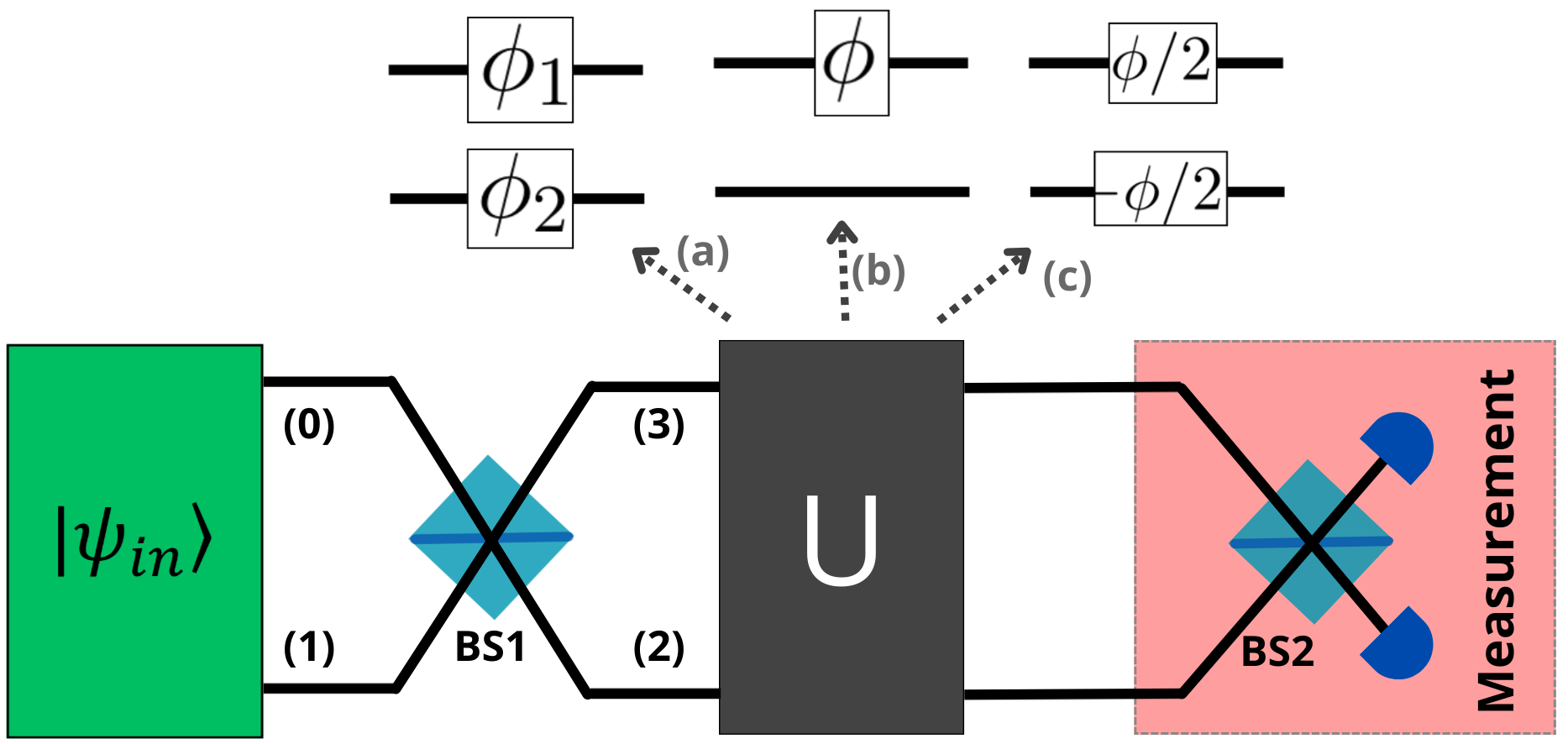} 
	\caption{This is a schematic setup of a MZI, which includes two independent phase shifts and two beam splitters (BS1 and BS2) with an adjustable transmission coefficient. The input state evolves through the interferometric paths before measurement.}\label{Fig1}
\end{figure}	

We begin by examining the most general case in which phase shifts $\phi_{1}$ and $\phi_{2}$ are applied to the upper and lower arms of the interferometer, respectively. To circumvent the issue of accounting for unavailable resources, such as an external phase reference, the two-parameter estimation method is employed, as discussed in the literature \cite{Lang2013, Takeoka2017}. When an external phase reference is absent, the quantity of interest becomes the phase difference between the two interferometer arms, defined as $\phi_{d} = \phi_{1} - \phi_{2}$. Consequently, it is convenient to represent the quantum Fisher information matrix (QFIM) in terms of the parameters $\phi_{s/d} = \phi_{1} \pm \phi_{2}$. This 2×2 matrix \cite{Liu2020, Abouelkhir2023(2), Abouelkhir2024} enables the joint estimation of $\phi_{s}$ and $\phi_{d}$ and is expressed as follows:
\begin{equation} \label{QFIM}
    \mathcal{F}=\left(\begin{array}{ll}
\mathcal{F}_{dd} & \mathcal{F}_{sd} \\
\mathcal{F}_{ds} & \mathcal{F}_{ss}
\end{array}\right)
\end{equation}
where
\begin{equation} \label{Fij}
    \mathcal{F}_{ij}=4\mathfrak{R}\{\langle\partial_{i}\psi|\partial_{j}\psi\rangle-\langle\partial_{i}\psi|\psi\rangle\langle\psi|\partial_{j}\psi\rangle\},
\end{equation}
here, $\mathfrak{R}$ represents the real, with the indices $i$ and $j$ associated with $\phi_{s}$ and $\phi_{d}$, respectively. Our focus is on the wavevector $|\psi\rangle$, which we denote as
\begin{equation}
    |\psi\rangle=\exp\left[{-i\frac{_{2}-\hat{m}_{3}}{2}\phi_{d}}\right]\exp\left[{-i\frac{\hat{m}_{2}+\hat{m}_{3}}{2}\phi_{s}}\right]|\psi_{23}\rangle,
\end{equation}
where $\hat{m}_{2}$ and $\hat{m}_{3}$ are the number operators corresponding to ports $2$ and $3$, respectively, with $\hat{m}=\hat{b}^{\dagger}\hat{b}$. To compute the elements of the QFIM, we need the transformations of the field operators. These transformations enable the representation of the output annihilation operators in terms of the input field operators, thus facilitating the calculation of the QFIM elements:
\begin{equation}
    \hat{b}_{2}=t\hat{b}_{0}+r\hat{b}_{1},\quad \hat{b}_{3}=r\hat{b}_{0}+t\hat{b}_{1}.
\end{equation}
The input field operators $\hat{b}_i$ and $\hat{b}_j^\dagger$ denote the annihilation and creation operators for mode $i$ and $j$, respectively, satisfying $[\hat{b}_i, \hat{b}_j^\dagger] = \delta_{ij}$, where $\delta_{ij}$ represents the Kronecker delta and $i,j= 0,1$. We have the equations
\begin{equation}
    |t|^2=1-|r|^2,
\end{equation}
and
\begin{equation}
    tr^{\ast}=-t^{\ast}r.
\end{equation}
These equations imply thatt $t^{\ast}r=\pm i|tr|$. Throughout this paper, we adopt the convention $t^{\ast}r=i|tr|$ without loss of generality and the following formula
\begin{equation}
    t=\cos{\frac{\kappa}{2}},\quad \mbox{and} \quad r=i\sin{\frac{\kappa}{2}}.
\end{equation}
We can rewrite the input-output field operator transformations as
\begin{align}  \label{transformations1}
    \hat{b}_{2}=&\cos{\frac{\kappa}{2}}\hat{b}_{0}+i\sin{\frac{\kappa}{2}}\hat{b}_{1},\\
    \hat{b}_{3}=&i\sin{\frac{\kappa}{2}}\hat{b}_{0}+\cos{\frac{\kappa}{2}}\hat{b}_{1}.
\end{align} 
From the QFIM, we obtain the QCRB matrix inequality, which is a lower bound on the variance of unbiased estimators in quantum parameter estimation. The QCRB is given by the inequality
\begin{equation}\label{QCRB}
Cov(\hat{\phi})\geq \frac{1}{N}\mathcal{F}^{-1},
\end{equation}
where $N$ is the number of repeated experiments, $\mathcal{F}^{-1}$ is the inverse of the QFIM given in eq. (\ref{QFIM}) and $Cov(\hat{\phi})$ is the covariance matrix of the unbiased estimator $\hat{\phi}$, including both $\phi_{d}$ and $\phi_{s}$, whose elements are 
\begin{equation}
    Cov(\hat{\phi})_{ij}=E(\hat{\phi}_{i}\hat{\phi}_{j})-E(\hat{\phi}_{i})E(\hat{\phi}_{j}),
\end{equation}
with E denotes the mathematical expectation. For the remainder of the paper, we consider $N = 1$. In particular, when we consider the sensitivity of the phase difference, we have
\begin{equation}
    (\Delta \phi_{d})^2\geq (\mathcal{F}^{-1})_{dd}.
\end{equation}
To avoid repetition of the matrix element $(\mathcal{F}^{-1})_{dd}$ of the inverse of matrix $\mathcal{F}^{-1}$, we introduce the two-parameter QFI
\begin{equation} \label{F(a)}
    \mathcal{F}^{(a)}=\frac{1}{(\mathcal{F}^{-1})_{dd}}=\mathcal{\mathcal{F}}_{dd}-\frac{(F_{sd})^2}{\mathcal{F}_{ss}}.
\end{equation}
In this way, the inequality (\ref{QCRB}) can be saturated, which implies that the two-parameter QCRB
\begin{equation}
\Delta \phi_{QCRB}^{(a)}= \frac{1}{\sqrt{\mathcal{F}^{(a)}}}.
\end{equation}
The components of the QFIM, specifically $\mathcal{F}_{ss}$, $\mathcal{F}_{dd}$, and $\mathcal{F}_{sd}$ are obtained by applying the definition given in equation (\ref{Fij}); 

{\small
\begin{align} 
\mathcal{F}_{ss}=&\Delta^2\hat{m}_{0}+\Delta^2\hat{m}_{1},\\
\mathcal{F}_{dd}=&\cos{\kappa}^2\left(\Delta^2\hat{m}_{0}+\Delta^2\hat{m}_{1}\right)+2\sin^2{\kappa}\left(
\langle\hat{m}_{0}\rangle\langle\hat{m}_{1}\rangle-|\langle\hat{b}_{0}\rangle|^2|\langle\hat{b}_{1}\rangle|^2-\mathfrak{R}\left\{\langle(\hat{b}_{0}^{\dagger})^2\rangle\langle\hat{b}_{1}^2\rangle-\langle\hat{b}_{0}^{\dagger}\rangle^2\langle\hat{b}_{1}\rangle^2\right\}\right)\\
&+\sin^2{\kappa}\left(\langle\hat{m}_{0}\rangle+\langle\hat{m}_{1}\rangle\right)-2|\sin{\kappa}|\cos{\kappa}\left(\mathfrak{I}\left\{\left(\langle\hat{b}_{0}^{\dagger}\hat{m}_{0}\rangle-\langle\hat{b}_{0}^{\dagger}\rangle\langle\hat{m}_{0}\rangle\right)\langle\hat{b}_{1}\rangle+\langle\hat{b}_{0}\rangle\left(\langle\hat{b}_{1}^{\dagger}\hat{m}_{1}\rangle-\langle\hat{m}_{1}\rangle\langle\hat{b}_{1}^{\dagger}\rangle\right)\right\}\right),\\
\mathcal{F}_{sd}=&\cos{\kappa}\left(\Delta^2\hat{m}_{0}-\Delta^2\hat{m}_{1}\right)+2|\sin{\kappa}|\mathfrak{I}\left\{\langle\hat{b}_{0}\rangle\langle\hat{b}^{\dagger}_{1}\rangle-\left(\langle\hat{m}_{0}\hat{b}_{0}\rangle-\langle\hat{m}_{0}\rangle\langle\hat{b}_{0}\rangle\right)\langle\hat{b}^{\dagger}_{1}\rangle+\langle\hat{b}_{0}\rangle\left(\langle\hat{b}_{1}^{\dagger}\hat{m}_{1}\rangle-\langle\hat{b}^{\dagger}_{1}\rangle\langle\hat{m}_{1}\rangle\right)\right\},
\end{align}
}
where $\Delta^{2}\hat{m}$ denotes the variance, defined as $\Delta^{2}\hat{m}=\langle\hat{m}^{2}\rangle-\langle\hat{m}\rangle^{2}$.\par

When a phase shift is applied to a single arm, specifically at output 3 of BS1, as indicated in Figure 1, the quantum state $|\psi\rangle$ takes the form $e^{-i\phi\hat{m}_{3}}|\psi_{23}\rangle$. Based on the definition in equation (\ref{Fij}), the corresponding single-parameter QFI denoted $\mathcal{F}^{(b)}$, is given by 
\begin{equation}
   \mathcal{F}^{(b)}= 4\Delta^2\hat{m}_{3},
\end{equation}
implying a QCRB
\begin{equation}
\Delta \phi_{QCRB}^{(b)}= \frac{1}{\sqrt{\mathcal{F}^{(b)}}}.
\end{equation}
By applying the field operator transformations (\ref{transformations1}), we get
{\small
\begin{align} \label{F(b)} \nonumber
\mathcal{F}^{(b)}=&4\sin^4{\frac{\kappa}{2}}\Delta^2\hat{m}_{0}+4\cos^4{\frac{\kappa}{2}}\Delta^2\hat{m}_{1}+\frac{1}{4}\sin^4{\kappa}\left(\langle\hat{m}_{0}\rangle+\langle\hat{m}_{1}\rangle+2\left(\langle\hat{m}_{0}\rangle\langle\hat{m}_{1}\rangle-|\langle\hat{b}_{0}\rangle|^2|\langle\hat{b}_{1}\rangle|^2\right)\right)\\ \nonumber
&-2\sin^2{\kappa}\mathfrak{R}\left\{\langle\hat{b}^2_{0}\rangle\langle(\hat{b}^{\dagger}_{1})^2\rangle-\langle\hat{b}_{0}\rangle\langle\hat{b}^{\dagger}_{1}\rangle\right\}-4\sin{\kappa}\mathfrak{I}\left\{\langle\hat{b}_{0}\rangle\langle\hat{b}^{\dagger}_{1}\rangle\right\}-8|\sin{\kappa}|\sin^2{\frac{\kappa}{2}}\mathfrak{I}\left\{\left(\langle\hat{m}_{0}\hat{b}_{0}\rangle-\langle\hat{m}_{0}\rangle\langle\hat{b}_{0}\rangle\right)\langle\hat{b}^{\dagger}_{1}\rangle\right\}\\
&-8|\sin{\kappa}|\cos^2{\frac{\kappa}{2}}\mathfrak{I}\left\{\langle\hat{b}_{0}\rangle\left(\langle\hat{b}^{\dagger}_{0}\hat{m}_{1}\rangle-\langle\hat{m}_{1}\rangle\langle\hat{b}^{\dagger}_{1}\rangle\right)\right\}.
\end{align}
}
By comparing equation (\ref{F(b)}) with the QFIM components derived in the first estimation scenario, we observe that the single-parameter QFI $\mathcal{F}^{(b)}$ can be rewritten in terms of the QFIM elements as follows: 
\begin{equation}
\mathcal{F}^{(b)}=\mathcal{F}_{dd}+\mathcal{F}_{ss}-2\mathcal{F}_{sd}.
\end{equation}
When the off-diagonal and diagonal elements of the QFIM satisfy the condition $\mathcal{F}_{sd} = \mathcal{F}_{ss}$, the two-parameter QFI in Eq.~(\ref{F(a)}) coincides with its single-parameter counterpart, implying $\mathcal{F}^{(a)} = \mathcal{F}^{(b)}$. More generally, one can establish the inequality
\begin{equation}
    \mathcal{F}^{(b)} \ge \mathcal{F}^{(a)},
\end{equation}
confirming that the single-parameter QFI always provides an upper bound for the two-parameter scenario.\par

In the final scenario, which is essentially a single-parameter estimation problem similar to the second scenario, we assume a single phase shift at output 3 of the first beamsplitter, which we model by the unitary operation as $U(\phi) = e^{i\frac{\phi}{2}(\hat{m}_{2}-\hat{m}_{3})}$, so that the QFI is given by
\begin{equation}
   \mathcal{F}^{(c)}= \Delta^2\hat{m}_{2}+\Delta^2\hat{m}_{3},
\end{equation}
similarly in equation (\ref{F(b)}), we obtain
{\small
\begin{align} \label{F(c)}
\mathcal{F}^{(c)}=&\left(\cos^4{\frac{\kappa}{2}}+\sin^4{\frac{\kappa}{2}}\right)\left(\Delta^2\hat{m}_{0}+\Delta^2\hat{m}_{1}\right)+\frac{1}{2}\sin^2{\kappa}\left(\langle\hat{m}_{0}\rangle+\langle\hat{m}_{1}\rangle+2\left(\langle\hat{m}_{0}\rangle\langle\hat{m}_{1}\rangle-|\langle\hat{b}_{0}\rangle|^2|\langle\hat{b}_{1}\rangle|^2\right)\right)\\\nonumber
&-\frac{1}{2}\sin^2{\kappa}\left(\langle\hat{b}^{2}_{0}\rangle\langle(\hat{b}^{\dagger}_{1})^{2}\rangle+\langle(\hat{b}^{\dagger}_{0})^{2}\rangle\langle\hat{b}^{2}_{1}\rangle-\langle\hat{b}_{0}\rangle^2\langle\hat{b}^{\dagger}_{1}\rangle^2-\langle\hat{b}^{\dagger}_{0}\rangle^2\langle\hat{b}_{1}\rangle^2\right)+\sin{\kappa}\cos{\kappa}\left(\langle\hat{b}^{\dagger}_{0}\hat{m}_{0}\rangle-\langle\hat{b}^{\dagger}_{0}\rangle\langle\hat{m}_{0}\rangle\right)\langle\hat{b}_{1}\rangle\\\nonumber
&-\sin{\kappa}\cos{\kappa}\left(\langle\hat{m}_{0}\hat{b}_{0}\rangle-\langle\hat{m}_{0}\rangle\langle\hat{b}_{0}\rangle\right)\langle\hat{b}^{\dagger}_{1}\rangle+\sin{\kappa}\cos{\kappa}\langle\hat{b}_{0}\rangle\left(\langle\hat{b}^{\dagger}_{1}\hat{m}_{1}\rangle-\langle\hat{b}^{\dagger}_{1}\rangle\langle\hat{m}_{1}\rangle\right)-\sin{\kappa}\cos{\kappa}\langle\hat{b}^{\dagger}_{0}\rangle\left(\langle\hat{m}_{1}\hat{b}_{1}\rangle-\langle\hat{m}_{1}\rangle\langle\hat{b}_{1}\rangle\right).\nonumber
\end{align}
}
and which implies the QCRB
\begin{equation}
\Delta \phi_{QCRB}^{(c)}= \frac{1}{\sqrt{\mathcal{F}^{(c)}}}.
\end{equation}

Following the general derivation of the quantum Fisher information in the Mach-Zehnder interferometric setup, we now apply these results to the case where the input consists of a generalized coherent state combined with a vacuum state. The input state is given as:
\begin{equation}\label{input state}
    |\psi_{in}\rangle=|\zeta\rangle_{1}\otimes|0\rangle_{0},
\end{equation}
where the subscript $i=GHA$ or $SU$ corresponds to GHA coherent state (Eq. \ref{z GHA}) and generalized SU(1,1) coherent state (Eq. \ref{z su}), respectively. The QFI is denoted by $\mathcal{F}_{i}$. Based on the previously derived expressions for the three types of QFIs, namely $\mathcal{F}^{(a)}$, $\mathcal{F}^{(b)}$, and $\mathcal{F}^{(c)}$, provided in Eqs. (\ref{F(a)}), (\ref{F(b)}), and (\ref{F(c)}), the corresponding QFIs for our input states can be directly obtained as follows:

\begin{align}
\mathcal{F}^{(a)}_{i}=&\sin^2{\kappa}\langle\hat{m}_{1}\rangle_{i},\\
\mathcal{F}^{(b)}_{i}=&4\cos^4{\frac{\kappa}{2}}\Delta^{2}\hat{m}_{1}+\sin^2{\kappa}\langle\hat{m}_{1}\rangle_{i},\\
\mathcal{F}^{(c)}_{i}=&(\cos^4{\frac{\kappa}{2}}+\sin^4{\frac{\kappa}{2}})\Delta^{2}\hat{m}_{1}+\frac{1}{2}\sin^2{\kappa}\langle\hat{m}_{1}\rangle_{i}.
\end{align}
In the scenario where the GHZ coherent state is combined with a vacuum state as input states, we compute the QFIs $\mathcal{F}_{GHZ}$ and obtain their analytical expressions:

{\small
\begin{align}
\mathcal{F}^{(a)}_{GHA}=&\sin^2{\kappa}N(|\zeta|)^{2}\sum_{m=0}^{\infty}\frac{|\zeta|^{2m}m}{(N_{m-1}!)^{2}},\\
\mathcal{F}^{(b)}_{GHA}=&4\cos^4{\frac{\kappa}{2}}N(|\zeta|)^{2}\left(\sum_{m=0}^{\infty}\frac{|\zeta|^{2m}m^{2}}{(N_{m-1}!)^{2}}-N(|\zeta|)^{2}\left(\sum_{m=0}^{\infty}\frac{|\zeta|^{2m}m}{(N_{m-1}!)^{2}}\right)^{2}
\right)+\sin^2{\kappa}N(|\zeta|)^{2}\sum_{m=0}^{\infty}\frac{|\zeta|^{2m}m}{(N_{m-1}!)^{2}},\\
\mathcal{F}^{(c)}_{GHA}=&(\cos^4{\frac{\kappa}{2}}+\sin^4{\frac{\kappa}{2}})N(|\zeta|)^{2}\left(\sum_{m=0}^{\infty}\frac{|\zeta|^{2m}m^{2}}{(N_{m-1}!)^{2}}-N(|\zeta|)^{2}\left(\sum_{m=0}^{\infty}\frac{|\zeta|^{2m}m}{(N_{m-1}!)^{2}}\right)^{2}
\right)+\frac{1}{2}\sin^2{\kappa}N(|\zeta|)^{2}\sum_{m=0}^{\infty}\frac{|\zeta|^{2m}m}{(N_{m-1}!)^{2}}.
\end{align}
}

In the balanced case, the QFIs become:

{\small
\begin{align}
\mathcal{F}^{(a)}_{GHA}=&N(|\zeta|)^{2}\sum_{m=0}^{\infty}\frac{|\zeta|^{2m}m}{(N_{m-1}!)^{2}},\\
\mathcal{F}^{(b)}_{GHA}=&N(|\zeta|)^{2}\left(\sum_{m=0}^{\infty}\frac{|\zeta|^{2m}m^{2}}{(N_{m-1}!)^{2}}-N(|\zeta|)^{2}\left(\sum_{m=0}^{\infty}\frac{|\zeta|^{2m}m}{(N_{m-1}!)^{2}}\right)^{2}
\right)+N(|\zeta|)^{2}\sum_{m=0}^{\infty}\frac{|\zeta|^{2m}m}{(N_{m-1}!)^{2}},\\
\mathcal{F}^{(c)}_{GHA}=&\frac{1}{2}N(|\zeta|)^{2}\left(\sum_{m=0}^{\infty}\frac{|\zeta|^{2m}m^{2}}{(N_{m-1}!)^{2}}-N(|\zeta|)^{2}\left(\sum_{m=0}^{\infty}\frac{|\zeta|^{2m}m}{(N_{m-1}!)^{2}}\right)^{2}
\right)+\frac{1}{2}N(|\zeta|)^{2}\sum_{m=0}^{\infty}\frac{|\zeta|^{2m}m}{(N_{m-1}!)^{2}}.
\end{align}
}

Similar analytical expressions for the generalized SU(1,1) coherent state can be obtained by replacing the average photon number and its variance with the corresponding expressions computed for the state given in Eq. (\ref{z su}). Based on these derivations, the corresponding QFIs and QCRBs for the different phase estimation scenarios are summarized in Table \ref{tab1}.\par

\begin{table}[h!]
\centering
\resizebox{\textwidth}{!}{%
\begin{tabular}{|c|c|c|}
\hline
\textbf{Scenario} & \textbf{QFI} & \textbf{QCRB} \\
\hline
Two-parameter estimation (a) 
& $F^{(a)}_{i} = \sin^{2}\kappa \, \langle \hat{m}_{1} \rangle_{i}$ 
& $\Delta\varphi^{(a)}_{\text{QCRB}} = \dfrac{1}{\sqrt{F^{(a)}_{i}}}$ \\
\hline
Single-arm phase shift (b) 
& $F^{(b)}_{i} = 4\cos^{4}\!\dfrac{\kappa}{2}\,\Delta^{2}\hat{m}_{1} + \sin^{2}\kappa \, \langle \hat{m}_{1} \rangle_{i}$ 
& $\Delta\varphi^{(b)}_{\text{QCRB}} = \dfrac{1}{\sqrt{F^{(b)}_{i}}}$ \\
\hline
Symmetric phase shift (c) 
& $F^{(c)}_{i} = \Big(\cos^{4}\!\dfrac{\kappa}{2} + \sin^{4}\!\dfrac{\kappa}{2}\Big)\Delta^{2}\hat{m}_{1} 
+ \tfrac{1}{2}\sin^{2}\kappa \, \langle \hat{m}_{1} \rangle_{i}$ 
& $\Delta\varphi^{(c)}_{\text{QCRB}} = \dfrac{1}{\sqrt{F^{(c)}_{i}}}$ \\
\hline
\end{tabular}
}
\caption{Summary of the analytical results for the QFI and the associated QCRB for different phase estimation scenarios.}\label{tab1}
\end{table}

The behavior of the QFI in three different estimation scenarios namely, $F^{(a)}$, $F^{(b)}$ and $F^{(c)}$, is investigated as a function of the transmission coefficient of the first beam splitter. This analysis is carried out for both the generalized Heisenberg algebra coherent state and the generalized su(1,1) coherent state, using the parameter values $a=0.5$ and $d=2e=0.2$. For the GHA coherent state, the two-parameter QFI represented by $F^{(a)}_{GHA}$ achieves its maximum at the balanced beam splitter condition $|t|=|r| = 1/\sqrt{2}$, while it vanishes in the extreme cases where $|t|=0$ or $|t|=1$. The single-parameter QFI $F^{(b)}_{GHA}$ exhibits a linear increase with the transmission coefficient, starting from zero and reaching its peak value at full transmission. Meanwhile, $F^{(c)}$ corresponding to a symmetric phase shift, remains approximately constant regardless of the value of the transmission coefficient. A similar trend is observed when considering the generalized su(1,1) coherent state. The QFI $F^{(a)}_{su}$ again peaks at the balanced configuration and vanishes at the extremes. The single-parameter QFI $F^{(b)}_{su}$ shows a linear dependence on the transmission coefficient, while $F^{(c)}_{su}$ stays effectively constant throughout. Comparison of the two coherent states reveals that, across all scenarios, the GHA coherent state provides larger QFI values than the generalized su(1,1) coherent state, indicating superior performance in estimating the phase parameter.

\section{Phase Sensitivity and Detection Schemes with Generalized Coherent State Inputs}
\begin{figure}[h] 
    \centering
		\includegraphics[scale=0.25]{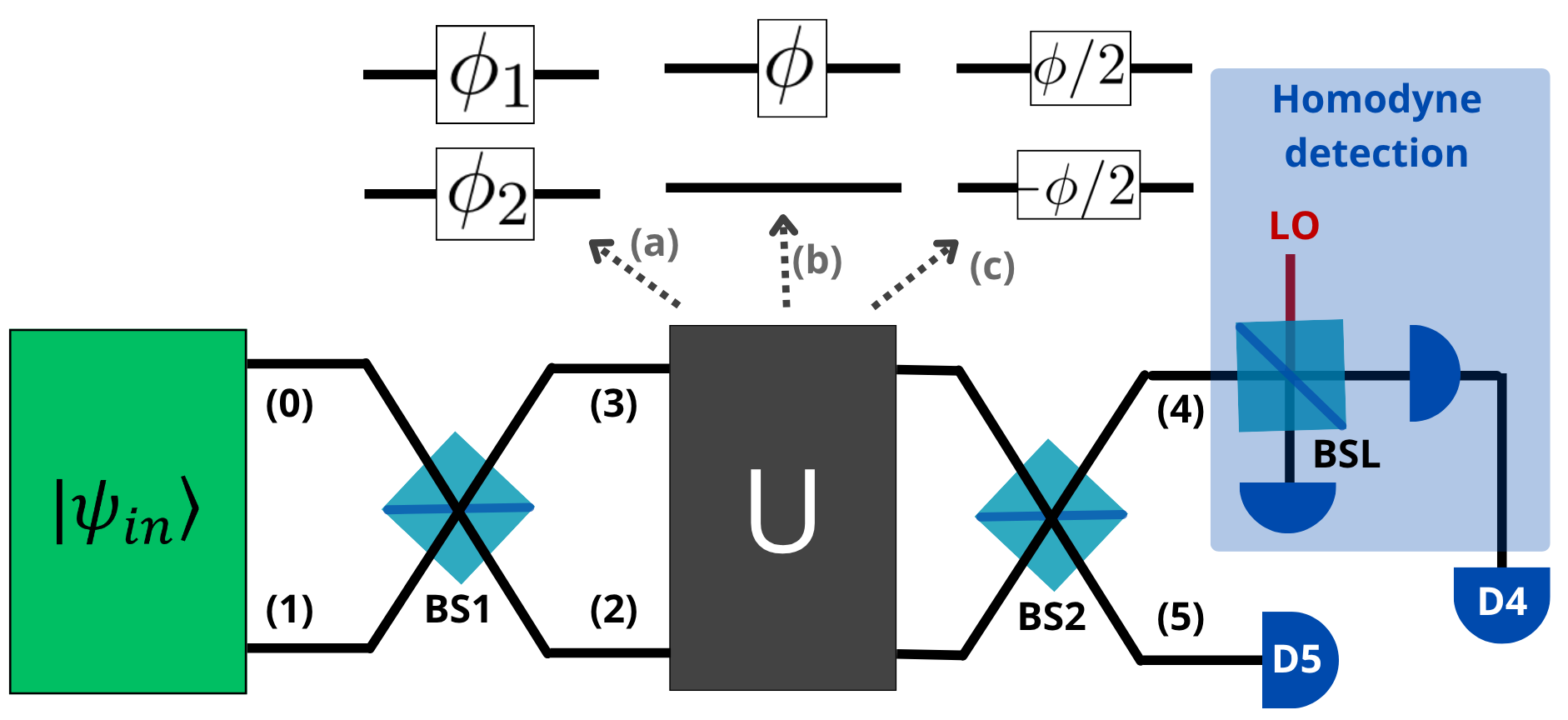} 
	\caption{This is a schematic of a Mach–Zehnder interferometer featuring three practical detection methods: intensity-difference detection, single-mode intensity detection and balanced homodyne detection. The input state, $|\psi_{\text{in}}\rangle$, evolves through a unitary transformation into the output state $|\psi_{\text{out}}\rangle$. The main objective is to estimate the phase shift $\theta$ between the interferometer arms using appropriate measurements.} \label{Fig2}
	\end{figure}

We now complete the Mach-Zehnder interferometer by introducing the second beam splitter (BS2), as depicted in Figure \ref{Fig2}, defined by its transmission ($t'$) and reflection ($r'$) coefficients. Subsequently, we analyze the performance of three practical detection strategies: difference intensity detection, single-mode intensity detection, and balanced homodyne detection.

We now turn to a brief discussion of the quantum parameter estimation problem. Let us consider a Hermitian operator $\hat{S}$, experimentally accessible, which depends on a parameter $\phi$. In our context, $\phi$ represents the phase shift introduced within a Mach–Zehnder interferometer and may not necessarily correspond to a directly observable quantity. The expectation value of this operator is expressed as

\begin{equation}
    \langle\hat{S}(\phi)\rangle=\langle\psi|\hat{S}(\phi)|\psi\rangle
\end{equation}

where the ket $|\psi\rangle$ denotes the quantum state (wave function) of the system. A small variation $\delta\phi$ in the parameter $\phi$ leads to a change in the expectation value, which can be approximated by the linear expansion
\begin{equation}
    \langle\hat{S}(\phi + \delta\phi)\rangle \approx \langle\hat{S}(\phi)\rangle + \frac{\partial \langle\hat{S}(\phi)\rangle}{\partial\phi} \delta\phi.
\end{equation}

The ability to experimentally distinguish the shift between $\langle\hat{S}(\phi + \delta\phi)\rangle$ and $\langle\hat{S}(\phi)\rangle$ is determined by the following criterion:
\begin{equation} \label{Condition}
    \langle\hat{S}(\phi + \delta\phi)\rangle - \langle\hat{S}(\phi)\rangle \geq \Delta\hat{S}(\phi),
\end{equation}
where $\Delta\hat{S}$ represents the standard deviation of the operator $\hat{S}$, defined as the square root of the variance $\Delta^2\hat{S}$:
\begin{equation}
    \Delta\hat{S} = \sqrt{ \langle \hat{S}^2 \rangle - \langle \hat{S} \rangle^2 }.
\end{equation}

If the inequality in Eq. (\ref{Condition}) is saturated by a particular value of $\delta\phi$, this minimal detectable variation defines the phase sensitivity, denoted by $\Delta\phi$ \cite{d’Ariano1994}, and it is given by:
\begin{equation}
    \Delta\phi = \frac{\Delta\hat{S}}{\left| \frac{\partial}{\partial\phi} \langle \hat{S} \rangle \right|}.
\end{equation}

In the following, $\phi$ represents the total phase shift inside the interferometer. This phase shift is divided into two parts: the first part, denoted as $\phi_{i}$, represents the quantity we want to measure. The second part is $\phi_{exp}$, which is experimentally controllable.
We have: $\phi=\phi_{i}+\phi_{exp}$.\par

In interferometry, the condition that the magnitude of $|\phi_{i}|$ is much smaller than that of $|\phi|$ is crucial because it means that the unknown phase shift $\phi_{i}$ has a limited effect on the total phase shift $\phi$. Therefore, the experimenter must adjust $\phi_{exp}$ to approach the optimal phase shift, denoted as $\phi_{opt}$, in order to achieve the best performance.\par

In what follows, we examine the phase sensitivity associated with each detection strategy under consideration namely, difference intensity detection, single-mode intensity detection, and balanced homodyne detection. The corresponding sensitivities are obtained as detailed below:

\subsection{Phase Sensitivity under Intensity-Difference Detection}

In the intensity-difference detection approach, which is exclusively sensitive to the relative phase shift $\phi = \phi_{1} - \phi_{2}$ between the two arms of the interferometer, we analyze the difference in output photocurrents measured at detectors D4 and D5 (see Figure~2). The corresponding output observable is defined as

\begin{equation}
    \hat{N}_{d} = \hat{b}^{\dagger}_{4} \hat{b}_{4} - \hat{b}^{\dagger}_{5} \hat{b}_{5}.
\end{equation}

To express $\hat{N}_{d}$ in terms of the input field operators, we first apply the transformation rules for the second beam splitter (BS2), whose transmission and reflection coefficients are given by $\cos{\frac{\kappa'}{2}}$ and $i\sin{\frac{\kappa'}{2}}$, respectively:

\begin{align} \label{transformations2}
    \hat{b}_{4} &= \cos{\frac{\kappa'}{2}}\,\hat{b}_{2} + i \sin{\frac{\kappa'}{2}}\,\hat{b}_{3}, \\
    \hat{b}_{5} &= i \sin{\frac{\kappa'}{2}}\,\hat{b}_{2} + \cos{\frac{\kappa'}{2}}\,\hat{b}_{3}.
\end{align}

By incorporating the initial beam splitter transformations (\ref{transformations1}), we obtain the full expressions for $\hat{b}_{4}$ and $\hat{b}_{5}$ in terms of the input modes $\hat{b}_0$ and $\hat{b}_1$:

\begin{align} \label{transformations22}
    \hat{b}_{4} &= \exp\left[{-i\phi_{2}}\right]\left[(\cos{\tfrac{\kappa}{2}}\cos{\tfrac{\kappa'}{2}} - \sin{\tfrac{\kappa}{2}}\sin{\tfrac{\kappa'}{2}} e^{-i\phi})\,\hat{b}_{0} + i(\cos{\tfrac{\kappa}{2}}\sin{\tfrac{\kappa'}{2}}e^{-i\phi} + \sin{\tfrac{\kappa}{2}}\cos{\tfrac{\kappa'}{2}})\,\hat{b}_{1}\right], \\
    \hat{b}_{5} &= \exp\left[{-i\phi_{2}}\right]\left[i(\cos{\tfrac{\kappa}{2}}\sin{\tfrac{\kappa'}{2}} + \sin{\tfrac{\kappa}{2}}\cos{\tfrac{\kappa'}{2}}e^{-i\phi})\,\hat{b}_{0} + (\cos{\tfrac{\kappa}{2}}\cos{\tfrac{\kappa'}{2}}e^{-i\phi} - \sin{\tfrac{\kappa}{2}}\sin{\tfrac{\kappa'}{2}})\,\hat{b}_{1}\right].
\end{align}

Substituting these into the expression for $\hat{N}_d$, we arrive at the following form:

\begin{align}
    \hat{N}_d &= \left[\left(\cos^2{\tfrac{\kappa}{2}} - \sin^2{\tfrac{\kappa}{2}}\right)\left(\cos^2{\tfrac{\kappa'}{2}} - \sin^2{\tfrac{\kappa'}{2}}\right) - \sin{\kappa} \sin{\kappa'} \cos{\phi} \right](\hat{m}_0 - \hat{m}_1) \nonumber \\
    &\quad + 2 \mathfrak{Re}\left\{ i\left[\sin{\kappa}(\sin^2{\tfrac{\kappa'}{2}} - \cos^2{\tfrac{\kappa'}{2}}) + (\sin^2{\tfrac{\kappa'}{2}}e^{-i\phi} - \cos^2{\tfrac{\kappa'}{2}}e^{i\phi})\sin{\kappa'} \right] \hat{m}_0 \hat{m}_1^{\dagger} \right\}.
\end{align}

The corresponding phase sensitivity is defined as

\begin{equation}
    \Delta\phi_{df} = \frac{\Delta \hat{N}_{d}}{ \left| \frac{\partial}{\partial\phi} \langle \hat{N}_{d} \rangle \right| }.
\end{equation}

The derivative of the expectation value of $\hat{N}_d$ with respect to $\phi$ reads

\begin{align}
    \frac{\partial}{\partial\phi} \langle \hat{N}_{d} \rangle &= \sin{\kappa} \sin{\kappa'} \sin{\phi} \left(\langle \hat{m}_{0} \rangle - \langle \hat{m}_{1} \rangle \right) \nonumber \\
    &\quad + 2|\sin{\kappa'}| \, \mathfrak{Re} \left\{ \left(\sin^2{\tfrac{\kappa}{2}} e^{-i\phi} + \cos^2{\tfrac{\kappa}{2}} e^{i\phi} \right) \langle \hat{b}_0 \rangle \langle \hat{m}_1^{\dagger} \rangle \right\}.
\end{align}

The variance $\Delta^2 \hat{N}_{d}$ can be expressed as

\begin{align} \label{Delta Nd}
    \Delta^2 \hat{N}_{d} &= A_{d}^{2} \left( \Delta^2 \hat{m}_{0} + \Delta^2 \hat{m}_{1} \right) + 4 A_{d} \, \mathfrak{Re} \left\{ C_{d} \left[ (\langle \hat{m}_0 \hat{b}_0 \rangle - \langle \hat{m}_0 \rangle \langle \hat{b}_0 \rangle) \langle \hat{b}_1^{\dagger} \rangle - \langle \hat{b}_0 \rangle (\langle \hat{b}_1^{\dagger} \hat{m}_1 \rangle - \langle \hat{m}_1 \rangle \langle \hat{b}_1^{\dagger} \rangle) \right] \right\} \\
    &\quad + |C_d|^2 \left( \langle \hat{m}_0 \rangle + \langle \hat{m}_1 \rangle \right) + 2 \mathfrak{Re} \left\{ C_d^2 \langle \hat{b}_0^2 \rangle \langle (\hat{b}_1^{\dagger})^2 \rangle - \langle \hat{b}_0 \rangle^2 \langle \hat{b}_1^{\dagger} \rangle^2 \right\} \nonumber  + 2|C_d|^2 \left( \langle \hat{m}_0 \rangle \langle \hat{m}_1 \rangle - |\langle \hat{b}_0 \rangle|^2 |\langle \hat{b}_1 \rangle|^2 \right).
\end{align}

Here, the coefficients $A_d$ and $C_d$ are given by

{\small
\begin{align} \label{Ad and Cd}
    A_{d} &= 1 - 2\sin^2{\left( \tfrac{\kappa + \kappa'}{2} \right)} + \sin{\kappa} \sin{\kappa'} (1 - \cos{\phi}), \\
    C_{d} &= |\sin{\kappa'}| \sin{\phi} + i\left[ |\sin{\kappa}|(1 - 2\cos^2{\tfrac{\kappa'}{2}}) + (1 - 2\cos^2{\tfrac{\kappa}{2}})|\sin{\kappa'}| \cos{\phi} \right],
\end{align}
}

and satisfy the normalization condition

\begin{equation}
    A_{d}^{2} + |C_{d}|^{2} = 1.
\end{equation}

It is worth noting that, within this detection scheme, the resulting phase sensitivity $\Delta\phi_{df}$ remains unchanged across both estimation configurations. Therefore, for simplicity, we adopt the unified notation $\Delta\phi_{df}$ to denote the phase sensitivity for this scheme.

\subsection{Phase Estimation via Single-Mode Intensity Detection}
In the single-mode intensity detection scheme, we restrict our attention to the measurement of a single output mode specifically, the photocurrent collected at output port 4. The corresponding observable is the photon number operator $\hat{m}_{4} = \hat{b}_{4}^{\dagger} \hat{b}_{4}$. The phase sensitivity in this context is quantified by
\begin{equation}
    \Delta\phi_{sing}=\frac{\Delta\hat{m}_{4}}{|\frac{\partial}{\partial\phi}\langle\hat{m}_{4}\rangle|}.
\end{equation}
Using the field operator transformation derived in Eq.~(\ref{transformations22}), we can express the expectation value of $\hat{m}_4$ in terms of the input fields:

\begin{align}
    \langle\hat{m}_{4}\rangle=&\left(\cos^{2}{\frac{\kappa}{2}}\cos^{2}{\frac{\kappa'}{2}}+\sin^{2}{\frac{\kappa}{2}}\sin^{2}{\frac{\kappa'}{2}}-\frac{1}{2}\sin{\kappa}\sin{\kappa'}\cos\phi\right)\langle\hat{m}_{0}\rangle\\ \nonumber
    &+\left(\cos^{2}{\frac{\kappa}{2}}\sin^{2}{\frac{\kappa'}{2}}+\cos^{2}{\frac{\kappa'}{2}}\sin^{2}{\frac{\kappa}{2}}+\frac{1}{2}\sin{\kappa}\sin{\kappa'}\cos\phi\right)\langle\hat{m}_{1}\rangle\\ \nonumber
    &+\frac{i}{2}\mathfrak{R}\left\{\left(\sin{\kappa}(2\cos^{2}{\frac{\kappa'}{2}}-1)+\sin{\kappa'}(\cos^{2}{\frac{\kappa}{2}}e^{-i\phi}-\sin^{2}{\frac{\kappa}{2}}e^{i\phi})\right)\langle\hat{b}_{0}^{\dagger}\rangle\langle\hat{b}_{1}\rangle\right\}.
\end{align}

Using the above equation, we immediately get
\begin{align}
\frac{\partial\langle\hat{m}_{4}\rangle}{\partial\phi}=&\frac{1}{2}\sin{\kappa}\sin{\kappa'}\sin\phi(\langle\hat{m}_{0}\rangle-\langle\hat{m}_{1}\rangle)+|\sin{\kappa'}|\mathfrak{R}\left\{\left(\cos^{2}{\frac{\kappa}{2}}e^{-i\phi}+\sin^{2}{\frac{\kappa}{2}}e^{i\phi}\right)\langle\hat{b}_{0}^{\dagger}\rangle\langle\hat{b}_{1}\rangle\right\}.\nonumber
\end{align}

To find $\Delta^{2}\hat{m}_{4}$, we first calculate the square of the operator $\hat{m}_{4}$. Then we get the final expression for $\Delta^{2}\hat{m}_{4}$ as
{\small\begin{align}\nonumber \label{Delta n4}
\Delta^{2}\hat{m}_{4}=&A_{0}^{2}\Delta^{2}\hat{m}_{0}+A_{1}^{2}\Delta^{2}\hat{m}_{1}+2|A_{01}^{2}|\mathfrak{R}\left\{\langle\hat{b}_{0}^{2}\rangle\langle(\hat{b}_{1}^{\dagger})^{2}\rangle-\langle\hat{b}_{0}\rangle^{2}\langle\hat{b}_{1}^{\dagger}\rangle^{2}\right\}+|A_{01}|^{2}\left(\langle\hat{m}_{0}\rangle+\langle\hat{m}_{1}\rangle+2\langle\hat{m}_{0}\rangle\langle\hat{m}_{1}\rangle-2|\langle\hat{b}_{0}\rangle|^{2}|\langle\hat{b}_{1}\rangle|^{2}\right)\\ 
&+2A_{0}\mathfrak{R}\left\{A_{01}(\langle\hat{m}_{0}\hat{b}_{0}^{\dagger}\rangle+\langle\hat{b}_{0}^{\dagger}\hat{m}_{0}\rangle-2\langle\hat{m}_{0}\rangle\langle\hat{b}_{0}^{\dagger}\rangle)\langle\hat{b}_{1}\rangle\right\}+2A_{1}\mathfrak{R}\left\{A_{01}\langle\hat{b}_{0}^{\dagger}\rangle(\langle\hat{m}_{1}\hat{b}_{1}\rangle+\langle\hat{b}_{1}\hat{m}_{1}\rangle-2\langle\hat{m}_{1}\rangle\langle\hat{b}_{1}\rangle)\right\},
\end{align}}
where
\begin{align}
\label{A0}
A_{0}=&\cos^{2}{\frac{\kappa}{2}}\cos^{2}{\frac{\kappa'}{2}}+\sin^{2}{\frac{\kappa}{2}}\sin^{2}{\frac{\kappa'}{2}}-\frac{1}{2}\sin{\kappa}\sin{\kappa'}\cos\phi,\\ 
\label{A1}
A_{1}=&\cos^{2}{\frac{\kappa}{2}}\sin^{2}{\frac{\kappa'}{2}}+\cos^{2}{\frac{\kappa'}{2}}\sin^{2}{\frac{\kappa}{2}}-\frac{1}{2}\sin{\kappa}\sin{\kappa'}\cos\phi,\\ 
\label{A01}
A_{01}=&\frac{i}{2}\left[\sin{\kappa}(2\cos^{2}{\frac{\kappa'}{2}}-1)+\sin{\kappa'}(\cos^{2}{\frac{\kappa}{2}}e^{-i\phi}-\sin^{2}{\frac{\kappa}{2}}e^{i\phi})\right].
\end{align}

\subsection{Phase Sensitivity via Balanced Homodyne Detection}

We now turn our attention to the balanced homodyne detection scheme, performed at output port 4, as illustrated in Fig.~2. In this configuration, the observable of interest is the quadrature operator defined by
\begin{equation}
    \hat{X}_{\phi_L} = \mathfrak{Re} \left\{ e^{-i\phi_L} \hat{b}_4 \right\},
\end{equation}
where $\phi_L$ denotes the phase of the local oscillator associated with the coherent reference state $|\gamma\rangle = |\gamma| e^{i\phi_L}$, with $\gamma$ being a complex amplitude.

The corresponding phase sensitivity in this detection protocol is given by
\begin{equation}
    \Delta\phi_{hom} = \frac{ \sqrt{ \Delta^2 \hat{X}_{\phi_L} } }{ \left| \frac{\partial \langle \hat{X}_{\phi_L} \rangle }{ \partial \phi } \right| }.
\end{equation}

Using the field transformations established in Eq.~(\ref{transformations22}), the expectation value of the quadrature operator becomes
\begin{align}
\langle \hat{X}_{\phi_L} \rangle = \mathfrak{Re} \Big\{ e^{-i\phi_L} \Big[ &\left( \cos{\tfrac{\kappa}{2}} \cos{\tfrac{\kappa'}{2}} e^{-i\phi_2} - \sin{\tfrac{\kappa}{2}} \sin{\tfrac{\kappa'}{2}} e^{-i\phi_1} \right) \langle \hat{b}_0 \rangle \nonumber \\
&+ i \left( \cos{\tfrac{\kappa}{2}} \sin{\tfrac{\kappa'}{2}} e^{-i\phi_1} + \cos{\tfrac{\kappa'}{2}} \sin{\tfrac{\kappa}{2}} e^{-i\phi_2} \right) \langle \hat{b}_1 \rangle \Big] \Big\}.
\end{align}

The corresponding variance is evaluated as
\begin{align}
\Delta^2 \hat{X}_{\phi_L} = \frac{1}{4} &+ 2\,\mathfrak{Re} \left\{ A^2 \Delta^2 \hat{b}_0 + B^2 \Delta^2 \hat{b}_1 \right\} \nonumber \\
&+ 2|A|^2 \left( \langle \hat{m}_0 \rangle - |\langle \hat{b}_0 \rangle|^2 \right) + 2|B|^2 \left( \langle \hat{m}_1 \rangle - |\langle \hat{b}_1 \rangle|^2 \right),
\end{align}
where the coefficients $A$ and $B$ are defined by
\begin{align} \label{A and B}
A &= \frac{1}{2} e^{-i(\phi_L + \phi_2)} \left( \cos{\tfrac{\kappa}{2}} \cos{\tfrac{\kappa'}{2}} - \sin{\tfrac{\kappa}{2}} \sin{\tfrac{\kappa'}{2}} e^{-i\phi} \right), \\
B &= \frac{i}{2} e^{-i(\phi_L + \phi_2)} \left( \cos{\tfrac{\kappa}{2}} \sin{\tfrac{\kappa'}{2}} e^{-i\phi} + \cos{\tfrac{\kappa'}{2}} \sin{\tfrac{\kappa}{2}} \right).
\end{align}

In scenario (b) depicted in Fig.\ref{Fig1}, where $\phi_1 = \phi$ and $\phi_2 = 0$, the absolute value of the derivative of $\langle \hat{X}_{\phi_L} \rangle$ with respect to $\phi$ reduces to

\begin{equation} \label{di}
    \left| \frac{\partial \langle \hat{X}_{\phi_L} \rangle }{ \partial \phi } \right| = \left| \mathfrak{Re} \left\{ e^{-i(\phi_L + \phi)} \left( i \sin{\tfrac{\kappa}{2}} \langle \hat{b}_0 \rangle + \cos{\tfrac{\kappa}{2}} \langle \hat{b}_1 \rangle \right) \right\} \right| \cdot \left| \sin{\tfrac{\kappa'}{2}} \right|.
\end{equation}
In contrast, for scenario (ii), where $\phi_1 = -\phi_2 = \phi/2$, we find
\begin{align} \label{dii}
\left| \frac{\partial \langle \hat{X}_{\phi_L} \rangle }{ \partial \phi } \right| = \frac{1}{2} \Big| \mathfrak{Re} \Big\{ ie^{-i\phi_L} \Big[ &\left( \cos{\tfrac{\kappa}{2}} \cos{\tfrac{\kappa'}{2}} e^{i\phi/2} + \sin{\tfrac{\kappa}{2}} \sin{\tfrac{\kappa'}{2}} e^{-i\phi/2} \right) \langle \hat{b}_0 \rangle \nonumber \\
&+ i \left( \cos{\tfrac{\kappa'}{2}} \sin{\tfrac{\kappa}{2}} e^{i\phi/2} - \cos{\tfrac{\kappa}{2}} \sin{\tfrac{\kappa'}{2}} e^{-i\phi/2} \right) \langle \hat{b}_1 \rangle \Big] \Big\} \Big|.
\end{align}

We proceed to compare the phase sensitivities obtained from the three detection schemes under consideration: difference intensity detection, single-mode intensity detection, and balanced homodyne detection. These are evaluated for input states composed of generalized SU(1,1) coherent states constructed either from the GHA or the generalized SU(1,1) algebra in combination with the vacuum. The resulting sensitivities are then benchmarked against the respective QCRBs determined from the QFIs presented in the preceding section.\par

Based on the analytical expressions for the phase sensitivities derived earlier, namely $\Delta\phi_d$, $\Delta\phi_{sing}$, and $\Delta\phi_{hom}$, and using the input state defined in Eq.~(\ref{input state}), we find that the phase sensitivities corresponding to our input states can be readily obtained as follows:\par

For a difference-intensity detection scheme, we get the final analytical expression of the phase sensitivity for the two types of SU(1,1) CSs, as
\begin{equation}
    \Delta\phi_{d}^{i}=\frac{\Delta_{i}\hat{N}_{d}}{\sin{\kappa}\sin{\kappa'}|\sin\phi \langle\hat{m}_{1}\rangle_{i}|},
\end{equation}
where the index $i=GHA$ or $su$ distinguishes between the GHA and generalized SU(1,1) coherent states. The variance $\Delta_{i}\hat{N}_{d}$ is computed using the operator expression in Eq. (\ref{Delta Nd}), and the parameters $A_{d}$ and $C_{d}$ are defined in Eq. (\ref{A and B}). In both scenarios (i) and (ii), we have the same result of phase sensitivity in this detection scheme.\par

For a single-mode intensity detection scheme, we obtain the phase sensitivity in all considered scenarios as
\begin{equation}
    \Delta\phi_{sing}^{i}=\frac{2\Delta_{i}\hat{m}_{4}}{\sin{\kappa}\sin{\kappa'}|\sin\phi \langle\hat{m}_{1}\rangle_{i}|}.
\end{equation}
where the variance $\Delta_{i}\hat{m}_{4}$ is given by Eq. (\ref{Delta n4}), and the coefficients $A_{0}$, $A_{1}$ and $A_{01}$ are defined in Eqs. (\ref{A0}, \ref{A1}, and \ref{A01}), respectively.\par

For a  balanced homodyne detection scheme, we have
\begin{equation}
\Delta^{2}_{i}\hat{X}_{\phi_{L}}=\frac{1}{4}+2\mathfrak{R}\left\{B^{2}\Delta^{2}_{i}\hat{b}_{1}\right\}+2|B|^{2}(\langle\hat{m}_{1}\rangle_{i}-|\langle\hat{b}_{1}\rangle_{i}|^{2}), 
\end{equation}
In the case of scenario (b), where $\phi_{1}=\phi$ and $\phi_{2}=0$, and assuming $\phi_{L}=\varphi$, the variance of the operator $\hat{X}_{\phi_{L}}$ is given by

\begin{align}
    \Delta_{i}^{2}\hat{X}_{\phi_{L}}=&\frac{1}{4}-\frac{1}{2}\left(\cos^{2}{\frac{\kappa}{2}}\sin^{2}{\frac{\kappa'}{2}}\cos2\phi+\cos^{2}{\frac{\kappa'}{2}}\sin^{2}{\frac{\kappa}{2}}+\frac{1}{2}\sin{\kappa}\sin{\kappa'}\cos\phi\right)\left(	|\langle\hat{b}^{2}_{1}\rangle_{i}|-|\langle\hat{b}_{1}\rangle_{i}|^{2}\right)\notag\\
    &+\frac{1}{2}\left(\cos^{2}{\frac{\kappa}{2}}\sin^{2}{\frac{\kappa'}{2}}+\cos^{2}{\frac{\kappa'}{2}}\sin^{2}{\frac{\kappa}{2}}+\frac{1}{2}\sin{\kappa}\sin{\kappa'}\cos\phi\right)\left(	\langle\hat{m}_{1}\rangle_{i}-|\langle\hat{b}_{1}\rangle_{i}|^{2}\right).
\end{align}
From equation (\ref{di}), we get

\begin{equation}
    |\frac{\partial\langle\hat{X}_{\phi_{L}}\rangle}{\partial\phi}|=\cos{\frac{\kappa'}{2}}\sin{\frac{\kappa}{2}}|\cos(\phi) \langle\hat{b}_{1}\rangle_{i}|,
\end{equation}
Using these results, we can determine the phase sensitivity

\begin{align}
    \Delta\phi_{hom}^{(b)}=&\frac{\sqrt{ \Delta_{i}^{2}\hat{X}_{\phi_{L}}}}{\cos{\frac{\kappa'}{2}}\sin{\frac{\kappa}{2}}|\cos(\phi) \langle\hat{b}_{1}\rangle_{i}|}.
\end{align}
In the case of scenario (c), where $\phi_{1}=-\phi_{2}=\phi/2$, the above variance is given by

\begin{align}
    \Delta_{i}^{2}\hat{X}_{\phi_{L}}=&\frac{1}{4}-\frac{1}{2}\left((\cos^{2}{\frac{\kappa}{2}}\sin^{2}{\frac{\kappa'}{2}}+\cos^{2}{\frac{\kappa'}{2}}\sin^{2}{\frac{\kappa}{2}})\cos\phi+\frac{1}{2}\sin{\kappa}\sin{\kappa'}\right)\left(	|\langle\hat{b}^{2}_{1}\rangle_{i}|-|\langle\hat{b}_{1}\rangle_{i}|^{2}\right)\notag\\
    &+\frac{1}{2}\left(\cos^{2}{\frac{\kappa}{2}}\sin^{2}{\frac{\kappa'}{2}}+\cos^{2}{\frac{\kappa'}{2}}\sin^{2}{\frac{\kappa}{2}}+\frac{1}{2}\sin{\kappa}\sin{\kappa'}\cos\phi\right)\left(	\langle\hat{m}_{1}\rangle_{i}-|\langle\hat{b}_{1}\rangle_{i}|^{2}\right).
\end{align}
From equation (\ref{dii}) we get

\begin{equation}
    |\frac{\partial\langle\hat{X}_{\phi_{L}}\rangle}{\partial\phi}|=|\cos{\frac{\kappa'}{2}}\sin{\frac{\kappa}{2}}-\cos{\frac{\kappa}{2}}\sin{\frac{\kappa'}{2}}||\cos(\frac{\phi}{2}) \langle\hat{b}_{1}\rangle_{i}|.
\end{equation}
Using these results, as in the last scenario, we can determine the phase sensitivity
\begin{align}
    \Delta\phi_{hom}^{(c)}=&\frac{\sqrt{ \Delta_{i}^{2}\hat{X}_{\phi_{L}}}}{|\cos{\frac{\kappa'}{2}}\sin{\frac{\kappa}{2}}-\cos{\frac{\kappa}{2}}\sin{\frac{\kappa'}{2}}||\cos(\frac{\phi}{2}) \langle\hat{b}_{1}\rangle_{i}|}.
\end{align}

The analytical results for the phase sensitivities obtained with the three considered detection schemes are summarized in Table \ref{tab2}.\par

\begin{table}[h!]
\centering
\resizebox{\textwidth}{!}{%
\begin{tabular}{|c|c|}
\hline
\textbf{Detection Scheme} & \textbf{Phase Sensitivity} \\
\hline
Intensity-difference detection 
& $\Delta\varphi^{\text{df}}_{i} = \dfrac{\Delta \hat{N}_{d}}{\sin\kappa \sin\kappa' |\sin\varphi| \langle \hat{m}_{1} \rangle_{i}}$ \\
\hline
Single-mode intensity detection 
& $\Delta\varphi^{\text{sing}}_{i} = \dfrac{2\Delta \hat{m}_{4}}{\sin\kappa \sin\kappa' |\sin\varphi| \langle \hat{m}_{1} \rangle_{i}}$ \\
\hline
Balanced homodyne detection (b) 
& $\Delta\varphi^{(b)}_{\text{hom},i} = \dfrac{\sqrt{\Delta^{2} \hat{X}^{\varphi_{L}}_{i}}}{\cos\!\dfrac{\kappa'}{2}\,\sin\!\dfrac{\kappa}{2}\,|\cos(\varphi)\langle \hat{b}_{1}\rangle_{i}|}$ \\
\hline
Balanced homodyne detection (c) 
& $\Delta\varphi^{(c)}_{\text{hom},i} = \dfrac{\sqrt{\Delta^{2} \hat{X}^{\varphi_{L}}_{i}}}{\big|\cos\!\dfrac{\kappa'}{2}\,\sin\!\dfrac{\kappa}{2} - \cos\!\dfrac{\kappa}{2}\,\sin\!\dfrac{\kappa'}{2}\big| \, |\cos(\tfrac{\varphi}{2})\langle \hat{b}_{1}\rangle_{i}|}$ \\
\hline
\end{tabular}
}
\caption{Summary of the analytical results for the phase sensitivities obtained with different detection schemes.}\label{tab2}
\end{table}

Loss mechanisms within the interferometer, as well as those arising from interactions with the environment, are beyond the scope of this study. For a detailed treatment of such effects, we refer the interested reader to Refs.~\cite{Dorner2009}.\par

In the present work, we do, however, include a simplified model of non-unit photodetection efficiency. This effect is assumed to be identical for all detectors and is characterized by the parameter $\eta \leq 1$, with $\eta = 1$ corresponding to ideal (lossless) detection. To incorporate this imperfection, we model each detector as being preceded by a virtual beam splitter with transmissivity $\sqrt{\eta}$, which results in the following transformation of the detected field operators:

\begin{equation}
\hat{b}'_k = \sqrt{\eta} \hat{b}_k + \sqrt{1 - \eta} \hat{b}_v,
\end{equation}
where $\hat{b}_v$ denotes the vacuum mode. Consequently, the detected photon number operator satisfies:
\begin{align}
\langle \hat{m}'_k \rangle &= \eta \langle \hat{m}_k \rangle,\\
\Delta^2 \hat{m}'_k &= \eta^2 \Delta^2 \hat{m}_k + \eta(1 - \eta) \langle \hat{m}_k \rangle.
\end{align}

The modified phase sensitivity for the single-mode intensity detection scheme is as follows:
\begin{equation}
\Delta \phi'_{\mathrm{sing}} = \frac{\sqrt{\Delta^2 \hat{m}_4 + \frac{1 - \eta}{\eta} \langle \hat{m}_4 \rangle}}{|\partial_\phi \langle \hat{m}_4 \rangle|}.
\end{equation}
In the special case where $\Delta^2 \hat{m}_4 = \langle \hat{m}_4 \rangle$ (shot-noise limit), this reduces to:
\begin{equation}
\Delta \phi' = \frac{\Delta \hat{m}_4}{\sqrt{\eta} |\partial_\phi \langle \hat{m}_4 \rangle|}.
\end{equation}

In the difference-inte
nsity detection scheme, the variance becomes:
\begin{equation}
\Delta^2 \hat{N}'_d = \eta^2 \Delta^2 \hat{N}_d + \eta (1 - \eta)(\langle \hat{m}_4 \rangle + \langle \hat{m}_5 \rangle),
\end{equation}
and the corresponding phase sensitivity is:
\begin{equation}
\Delta \phi'_{\mathrm{d}} = \frac{\sqrt{\Delta^2 \hat{N}_d + \frac{1 - \eta}{\eta} (\langle \hat{m}_4 \rangle + \langle \hat{m}_5 \rangle)}}{|\partial_\phi \langle \hat{N}_d \rangle|}.
\end{equation}

Finally, for the balanced homodyne detection scheme, the modified phase sensitivity is given by:
\begin{equation}
\Delta \phi'_{\mathrm{hom}} = \frac{\sqrt{\Delta^2 \hat{X}_{\phi_L} + \frac{1}{4} \frac{1 - \eta}{\eta}}}{|\partial_\phi \langle \hat{X}_{\phi_L} \rangle|}.
\end{equation}

In all the above expressions, we observe that the inequality $\Delta \phi \leq \Delta \phi'$ is always satisfied, meaning that detection losses degrade the phase sensitivity. Among the three considered schemes, the balanced homodyne detection exhibits the highest robustness against non-unit detection efficiency, particularly around optimal working points.\par

\begin{figure}[H]
  \centering
  \begin{minipage}{0.48\textwidth}
    \centering
    \includegraphics[width=\linewidth]{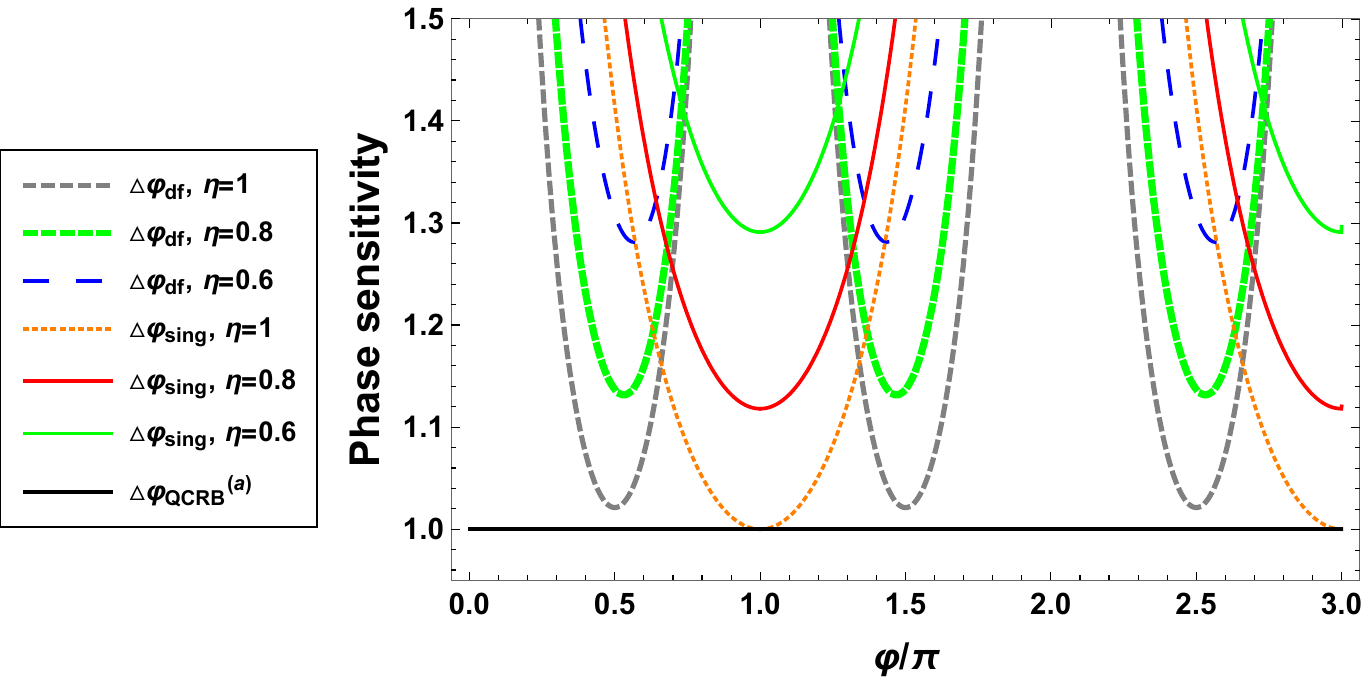}
    \subcaption{}
  \end{minipage}%
  \hfill
  \begin{minipage}{0.48\textwidth}
    \centering
    \includegraphics[width=\linewidth]{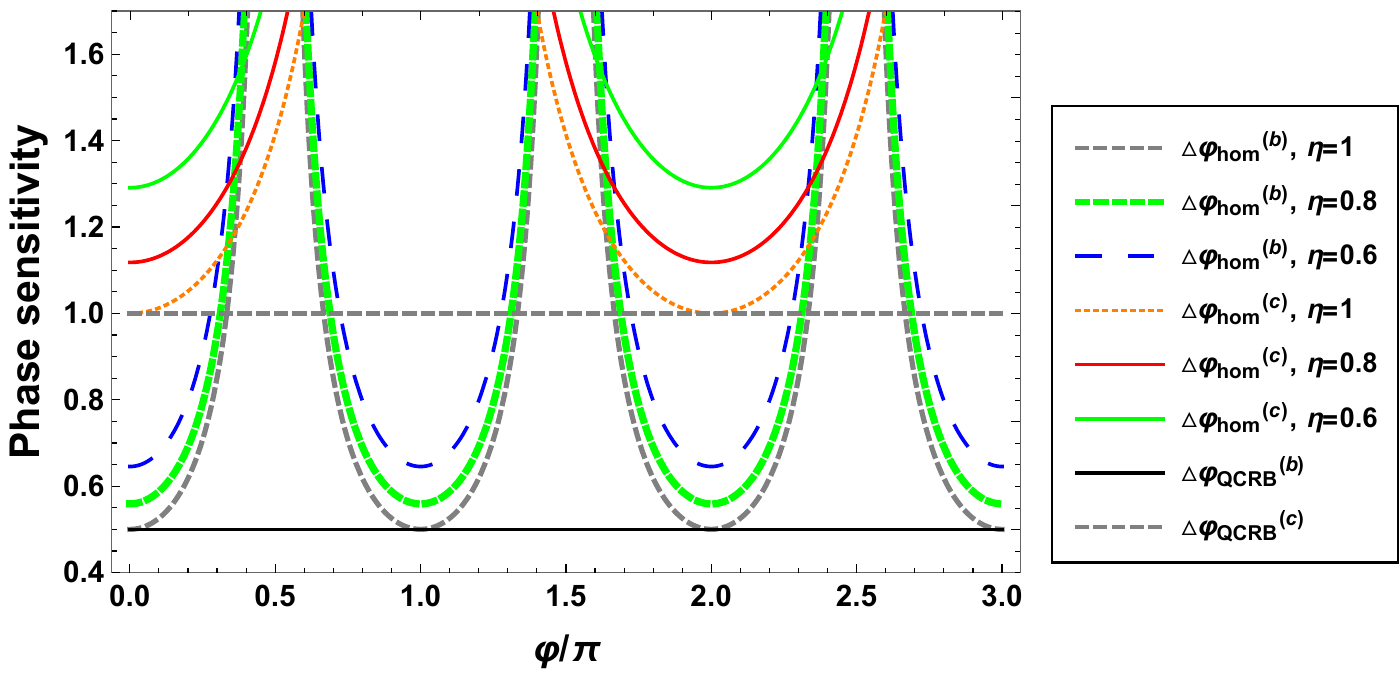}
    \subcaption{}
  \end{minipage}

  \vspace{1em}

  \begin{minipage}{0.48\textwidth}
    \centering
    \includegraphics[width=\linewidth]{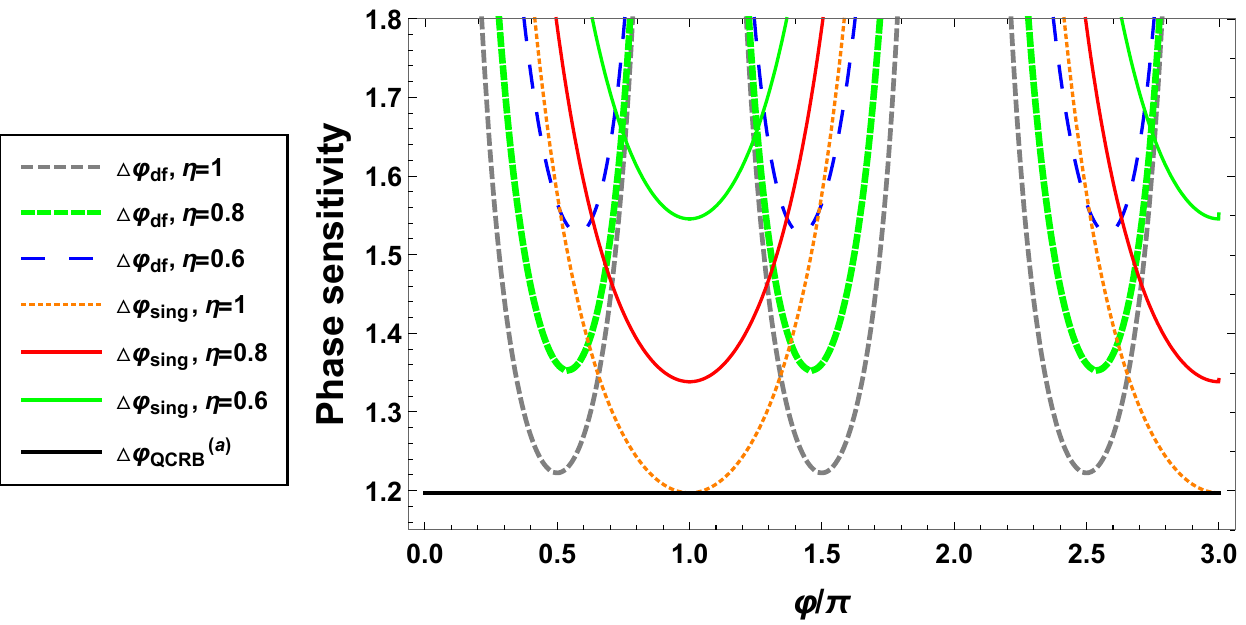}
    \subcaption{}
  \end{minipage}%
  \hfill
  \begin{minipage}{0.48\textwidth}
    \centering
    \includegraphics[width=\linewidth]{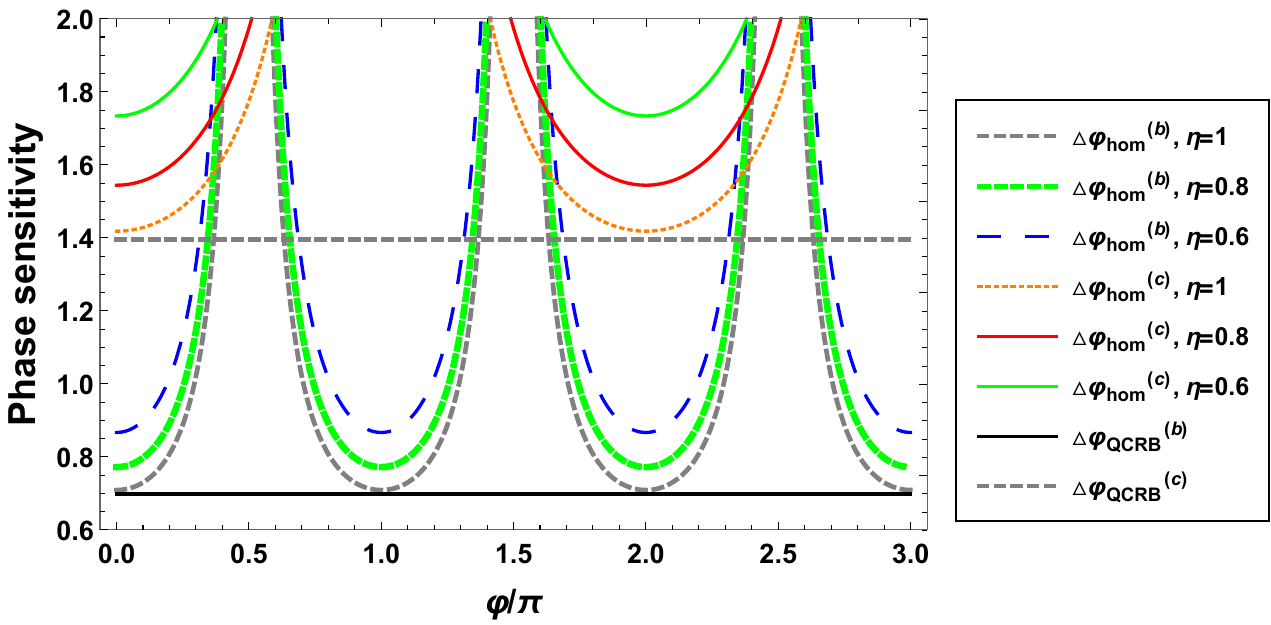}
    \subcaption{}
  \end{minipage}
\caption{Phase sensitivity $\Delta\phi$ as a function of the interferometric phase shift $\phi$, evaluated for three detection strategies: difference-intensity detection ($\Delta\phi_{\mathrm{df}}$), single-mode intensity detection ($\Delta\phi_{\mathrm{sing}}$), and balanced homodyne detection in two configurations ($\Delta\phi^{(b)}_{\mathrm{hom}}$ and $\Delta\phi^{(c)}_{\mathrm{hom}}$). Results are presented for two types of input states: (a) and (c) correspond to the GHA coherent state, while (b) and (d) correspond to the generalized SU(1,1) coherent state. The phase sensitivities are compared against the corresponding quantum Cramér Rao bounds: $\Delta\phi^{(a)}_{\mathrm{QCRB}}$, $\Delta\phi^{(b)}_{\mathrm{QCRB}}$, and $\Delta\phi^{(c)}_{\mathrm{QCRB}}$. The gray curve represents $\Delta\phi_{\mathrm{df}}$, the green curve $\Delta\phi_{\mathrm{sm}}$, the blue curve $\Delta\phi^{(b)}_{\mathrm{hom}}$, and the orange curve $\Delta\phi^{(c)}_{\mathrm{hom}}$. The QCRBs are indicated by red, green, and black lines, respectively.} \label{Fig3}
\end{figure}

Figure \ref{Fig3} presents the variation of the three phase sensitivities as functions of the phase shift in the MZI, for both the GHA coherent state and the generalized SU(1,1) coherent state, with parameters $k = 1$ and $v = 1$. These phase sensitivities are evaluated under three realistic detection strategies: difference-intensity detection, single-mode intensity detection, and balanced homodyne detection. The results are compared with the corresponding QCRBs obtained from the QFI.\par

For both the difference-intensity and single-mode intensity detection schemes, we adopt a balanced beam splitter configuration characterized by $t = 1/\sqrt{2}$ and $r = i/\sqrt{2}$, corresponding to the optimal setup in the ideal (lossless) case. For the balanced homodyne detection scheme, optimal phase sensitivity is achieved when the first beam splitter is fully transmissive ($t=1$) and the second beam splitter is fully reflective ($t'=0$). The red and blue curves in Figure \ref{Fig3} correspond to the phase sensitivities obtained using difference-intensity and single-mode detection schemes, respectively. Both schemes show optimal points where the phase sensitivities reach the QCRB, confirming the theoretical prediction based on the two-parameter QFI.\par

The dashed and solid red curves represent two configurations of balanced homodyne detection. As illustrated in the figure, these configurations provide phase sensitivities that closely approach the QCRB over a broad range of phase shifts. Notably, the phase sensitivity $\Delta\phi_{\mathrm{hom}}^{(b)}$ exhibits superior performance compared to $\Delta\phi_{\mathrm{df}}$, suggesting a metrological advantage when an external phase reference (i.e., local oscillator) is employed, regardless of the type of input state used. The overall input state, including the local oscillator, is given by
\begin{equation}
|\psi\rangle = |\psi_{\mathrm{in}}\rangle \otimes |\gamma\rangle = |\zeta_i, k\rangle \otimes ||\gamma| e^{i\phi_L}\rangle.
\end{equation}
Here, the interferometric configuration involves two arms: the signal arm traversing both beam splitters with intermediate phase shift $\phi_1$, and the local oscillator arm entering the balanced beam splitter of the homodyne setup. Previous works on q-deformed and non-linear coherent states have shown that deformation enhances squeezing and sub-Poissonian statistics, enabling phase sensitivities beyond the SNL \cite{Solomon1990, Wang2000}. Our contribution goes further by employing generalized Heisenberg and generalized su(1,1) algebras with multiple deformation parameters, offering greater tunability and control of non-classical properties. This richer algebraic structure optimizes quantum Fisher information and phase sensitivity, thereby extending and complementing previous deformed-algebra approaches in quantum metrology.\par

To explore the robustness of these schemes, we also investigate the impact of non-unit detection efficiency by comparing the ideal case $(\eta = 1)$ with a lossy scenario $(\eta = 0.6)$. As shown in Figure \ref{Fig3}, detection losses degrade all considered phase sensitivities, particularly around their optimal working points. Nevertheless, a general trend is observed: for any internal phase shift $\phi$, we have $\Delta\phi \leq \Delta\phi'$, confirming the expected sensitivity degradation due to imperfect detection.\par

Among the detection strategies, balanced homodyne detection proves to be the most resilient to losses. Unlike intensity-based schemes, whose performance drops significantly in the presence of inefficiencies, the homodyne scheme maintains a relatively stable sensitivity across a broad range of phase shifts. This robustness makes homodyne detection particularly attractive for experimental implementations where non-ideal detectors and optical losses are inevitable.\par

\begin{figure}[h] 
	\centering
  \begin{minipage}{0.48\textwidth}
    \centering
    \includegraphics[width=\linewidth]{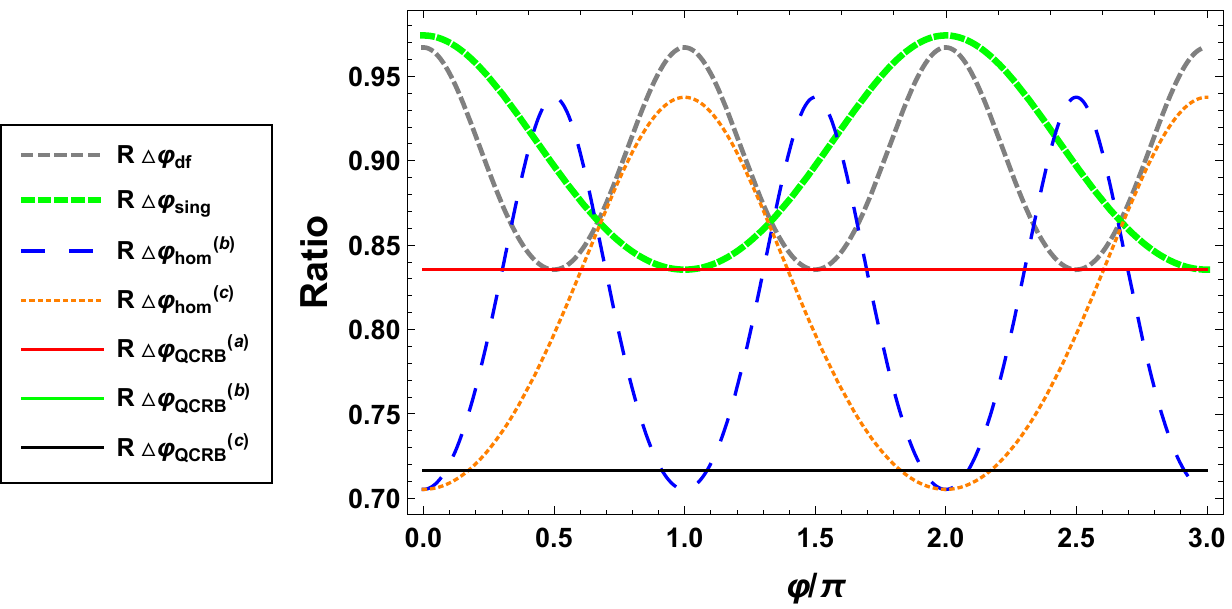}
    \subcaption{}
  \end{minipage}%
  \hfill
  \begin{minipage}{0.48\textwidth}
    \centering
    \includegraphics[width=\linewidth]{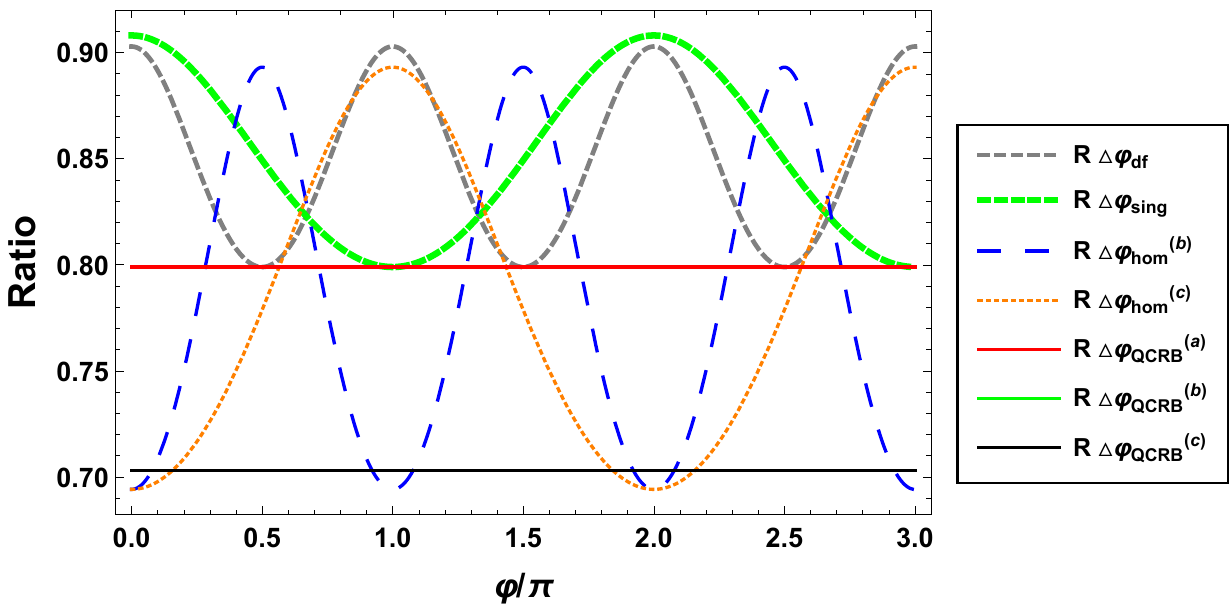}
    \subcaption{}
  \end{minipage}
    
	\caption{The performance ratio $R=\Delta\phi_{i}/\Delta\phi_{j}$, with $i$ representing the GHA coherent input state and $j$ representing the generalized SU(1,1) coherent input state, of optimal phase estimation in different detection schemes; panel (a) for $a=0.5$, panel (b) for $a=0.7$.} \label{Fig4}
	\end{figure}	

To compare the performance of optimal phase estimation in different detection schemes with the two types of input states, we use a technique where we introduce the ratio between the phase sensitivities of these two states in different detection schemes as $R=\Delta\phi_{i}/\Delta\phi_{j}$, where $i$ and $j$ refer to the GHA and generalized SU(1,1) coherent states, respectively. As a result, when the ratio $R<1$, the GHA coherent state yields a smaller phase uncertainty, indicating better performance. As shown in Figure \ref{Fig4}, the values of the phase sensitivities satisfy the inequality $\Delta\phi_{GHA}<\Delta\phi_{su}$ in all detection schemes considered, confirming that the GHA coherent states provide enhanced sensitivity and improved precision in the estimation of the phase shift compared to the generalized SU(1,1) coherent states.

\section{CONCLUSION}
Optimizing the phase sensitivity of a MZI hinges on the judicious selection of input states and detection schemes. The QFI plays a pivotal role in determining the fundamental limits of precision, enabling the identification of optimal operating conditions for phase estimation. In this study, we derived the QCRBs for both two-parameter and single-parameter estimation scenarios, focusing on generalized coherent states constructed from deformed \(su(1,1)\) and Heisenberg algebras. We evaluated the phase sensitivity of these states under three practical detection methods: difference intensity detection, single-mode intensity detection, and balanced homodyne detection, and compared their performance with the corresponding QCRBs.\par  

Our findings reveal that the GHA coherent states consistently outperform their generalized \(su(1,1)\) counterparts in terms of phase sensitivity across all detection schemes. Notably, balanced homodyne detection, when combined with an external phase reference, achieves phase sensitivities that closely approach the QCRB, particularly in configurations where the beam splitters are optimized (e.g., fully transmissive or reflective). This underscores the advantage of incorporating external references in quantum metrology to enhance measurement precision.\par

Beyond coherent states, entangled states hold significant promise for surpassing classical limits in quantum interferometry. While coherent states are optimal for classical measurements, entangled states can exploit quantum correlations to achieve sensitivities beyond the standard quantum limit. Future research could explore the integration of entanglement-enhanced strategies into interferometric setups, potentially unlocking new avenues for ultra-precise measurements in applications such as gravitational wave detection and environmental sensing.\par

In summary, this work highlights the flexibility and enhanced tunability of generalized coherent states for quantum metrology, while also pointing toward the potential of entangled states to further push the boundaries of precision in interferometric measurements. These insights pave the way for future experimental implementations and theoretical advancements in the field of quantum-enhanced sensing.

\end{document}